# Some geometrical aspects of gravitational waves using continuum mechanics analogy: State of the art and potential consequences

**David Izabel[1 3] Yves Remond[1] Matteo Luca Ruggiero[2]**

[1] University of Strasbourg, ICube Laboratory, CNRS, France
[2] University of Turin, Italy
[3] University of L'Aquila, Italy

E-mail: d.izabel@aliceadsl.fr

### Abstract

In this study, the authors employ the analogy between continuum mechanics and general relativity to investigate, from the perspective of elasticity and crystal plasticity, the deformations of space measured by LIGO/VIRGO interferometers during the passage of gravitational waves over Earth. The results of different innovative or existing mechanical models are compared with each other and compared with the observations in the framework of general relativity and Einstein-Cartan theory. Despite limitations, there is a convergence of results: the polarizations of gravitational waves can be viewed as expressions of an equivalent elastic media deformation tensor. Additionally, an anisotropy of space properties is unavoidable at the measurement point of the gravitational wave if we rely on the current first-order general relativity, which predict that gravitational waves generate deformations only in transverse planes. It is demonstrated that the classical polarizations of general relativity can be associated with a state of pure torsion in the analogous elastic medium and acted upon by the rotation of massive bodies such as black holes. This approach–involves a transverse isotropic medium composed of independent sheets that deform perpendicularly to the direction of propagation of these waves. Considering geometric torsion in general relativity, associated with plastic crystallography, allows for the examination of complementary polarizations in the direction of wave propagation. This makes it possible to connect these sheets and reconstruct a complete, coherent 3D environment.

## 1. Introduction

Einstein's theory of general relativity is over 100 years old and is now widely verified. Thus, according to this theory, space-time could be an elastic, deformable physical object. These distortions disappear when the object that created them disappears. Hence the notion of the elasticity of space. Gravitation is thus a manifestation of the geometric deformation of space-time under the effect of the masses or energy density found therein. The manifestations of the deformations of this space-time are now known and measured with great precision in several very specific situations. From a historical perspecpective, we can mention the variation in the apparent position of stars placed behind the sun during an eclipse measured by Edington in 1919 **[1]**, the expansion of the universe where galaxies are "fixed" in a space that expands in an increasingly accelerated manner characterized by Hubble's law established in 1929. More recently, it is relevant to remember the entrainment of the reference frame of space-time by angular distortion by the rotation of the Earth (experiment conducted with the satellite gravity prob B carried out with gyroscopes placed in orbit 642 km from the Earth from 2004 to 2010, Lense-Thirring (frame dragging) effects and geodetic precession) **[2]**, the simultaneous deformations of elongation and shortening in each of the arms of the LIGO/VIRGO interferometers during the passage of gravitational waves measured for the first time on 15 September 2014, **[3]**, **[4]**. A thorough overview of 100 years of testing general relativity can be found in **[45]**. All these manifestations of space-time distortions have led many physicists and mechanics researchers such as A Sakharov **[5]**, J.L. Synge **[6]**, C. B Rayner **[7]**, R. Grot **[8]**, TG. Tenev and M.F. Horstemeyer **[9]**, P.A. Millette **[10]**, D. Izabel **[11]**, **[12]**, T. Damour **[13]**, consider that the theory of general relativity in the weak field could be considered by analogy as a kind of theory of elasticity, a kind of Hooke's law of a deformable elastic space-time medium.

Thus, either some authors begin with general relativity and attempt to present it within the formalism of continuous mechanics, while other start with 3D continuum mechanics and generalize it to 4D by introducing a "mechanistic" metric that includes the effect of time.

The latest work that, to our knowledge, provides an updated assessment of this topic was giving by **[14**, **15**, **16]**, based on the following seminal works **[17-29]**.

Summarizing the theoretical context in which our work is situated, we will assume that both general relativity and continuum mechanics are based on the principle of general covariance, which, as B. Kolev aptly summarizes in **[15]**, requires the introduction of three components:

- Lagrangian-type functionals $\mathcal{L}$, depending on the metric g defined on the 4-dimensional universe manifold $\boldsymbol{M}$, and depending on various fields $\Psi$ ;
- Tensors, such as the energy-momentum tensor, dependent on these fields ;
- Field equations.

Our approach is therefore placed in the framework of the general covariance of field equations by the group of diffeomorphisms. Or expressed explicitly, a Lagrangian $\mathcal{L}(g,\Psi,..)$ is general covariant if it verifies $\mathcal{L}(\varphi^{*}g,\varphi^{*}\Psi)=\mathcal{L}(g,\psi)$ for any diffeomorphism $\varphi.$ Note, still with B. Kolev, that the functional $\mathfrak{H}(g)=\int R_g\,Vol_g,$ where $R_g$ represents the scalar curvature, is general covariant. Its gradient $L^2,$ the Einstein tensor $\boldsymbol{G}(g)$ is also general covariant. It consequently verifies: $\forall\varphi,\boldsymbol{G}_{\varphi^{*}g}=\varphi^{*}\boldsymbol{G}(g),$ and $div^{g}\boldsymbol{G}(g)=0.$ An energy-momentum tensor $T(g,\Psi)$ verifying the Einstein equation $\boldsymbol{G}(g)=T(g,\Psi)$ therefore verify the mechanics type equation: $div\,T=0.$ The rest of this work will therefore concern energy-momentum tensors verifying these two properties. The reader is referred to the work of J. M. Souriau **[27**, **28]** and his successors already cited, for the definitions of the perfect matter field (as a section of a vector bundle) and the conformations, allowing the proposal of relativistic constitutive laws **[16]**.

It is within the framework of this analogy that we mainly place ourselves in this paper. We will not develop further the very theoretical aspects described above. Nor will we address the controversies and debates that these concepts continue to generate, but we will concentrate on the consequences on the mechanical properties of the medium that the experiments induce. Following these preliminary remarks, we can define this “elastic gravitational analogy” based on the three principles of equivalences.

- The perturbations $h_{ij}$ of space in the presence of gravitational waves are linked to the Green-Lagrange covariant tensor: $D = \frac{1}{2}(\varphi^* g - g)$, which will be assimilated in the following, under infinitesimal deformations, as the geometric linearization of the strain tensor $D = \varepsilon = \varepsilon_{ij}$,
- Einstein's equation connects, within the medium, the strain tensor (versus metric perturbation in weak field) to the equivalent stress field, akin to Hooke's law, with the aid of an equivalent compliance matrix.
- The energy density of space-time itself $\rho c^2$ is correlated, through quantum field theory or the Casimir effect, with a non-zero energy density of the vacuum. This vacuum energy density is mirrored in the analogy by the Young's modulus Y of the equivalent elastic medium, as given by the equation $Y = \rho c^2$.

In addition to these hypotheses, we must recall the debates on the relativistic equivalent validity of the stress field used in continuum mechanics **[30]**. Our approach adds some further elements to this debate. However, when deformations occur in a vacuum far from the spacetime loading, as is the case with gravitational waves arriving on Earth, the stress-energy tensor is $T = T_{\mu\nu} = 0$.

To maintain a complete Hooke's law, it is necessary to consider an elastic strain energy tensor of the vacuum itself: $T_{e,\mu\nu}$, linked to the deformations correlated with $h_{\mu\nu}$ (the small perturbation of the Minkowski's tensor). This is one way to calculate the gravitational wave energy or to study the spacetime itself as an equivalent elastic medium **[60]**, **[61]**, **[62]**.

We add, however, that continuum mechanics models are based on a primary concept, which is the kinematics of the phenomena. The concept of stress field is a secondary concept which is introduced only after the kinematics choice using the virtual power principle **[31-32]**, named virtual power theorem if we consider as primary principle, the principle of general covariance. The constitutive laws which link these two concepts, or their time derivatives, are well defined by the Local State Method **[33]**.

## 2. Methods

The following methodology was employed to evaluate some geometrical aspects of gravitational waves in linearized general relativity from the perspective of the analogy with continuum mechanics.

1) Investigation of the discrepancy between the deformations measured during the passage of gravitational waves and the theoritical predictions of linearized general relativity, using deformation measurements made by LIGO/VIRGO interferometers,

2) Analysis of test masses arranged in a circle of the different deformations and associated polarizations according to various versions of general relativity (classical in the first-order, second-order in gravito-electromagnetism, modified with Einstein-Cartan torsion, and other modified versions),

3) Study of space deformations during the passage of a gravitational wave by considering either existing or new models, such as interferometer arms, torsional space cylinders, isotropic transverse media of elastic solids,

4) Comparison of the results from classical first- and second-order or modified theories of general relativity (Einstein-Cartan and others) with the predictions of various continuum mechanics. Specifically, we study the potential complementary deformations of the equivalent cosmic medium and their consequences on the characteristics of the media. We also explore interactions with the theory of defects in crystalline media regarding the number and types of gravitational wave polarizations.

## 3. Gravitational waves in Relativistic Theories of Gravity

### *3.1 General Relativity*

According to Einstein's theory of general relativity, gravitation is the geometry of spacetime: specifically, spacetime is a four-dimensional pseudo-riemannian manifold $\boldsymbol{M}$, which is a pair $(\boldsymbol{M}, \boldsymbol{g}_{\mu\nu})$, where $\boldsymbol{M}$ is a connected 4-dimensional Haussdorf manifold and $\boldsymbol{g}_{\mu\nu}$ is the metric tensor[1]. Due to its Riemannian structure, spacetime is endowed with an affine connection compatible with the metric, known as the Levi-Civita connection. We are interested in gravitational waves, which are particular vacuum solutions of Einstein's equations:

$$G^{\mu\nu} = -\frac{8\pi G}{c^4} T^{\mu\nu} \quad (1)$$

where $\boldsymbol{G}^{\mu\nu}$ is the Einstein tensor defined in terms of the Ricci tensor $\boldsymbol{R}^{\mu\nu}$ and scalar curvature R, as $\boldsymbol{G}^{\mu\nu} = \boldsymbol{R}^{\mu\nu} - \frac{1}{2}\boldsymbol{g}^{\mu\nu}\mathbf{R}$ and $\boldsymbol{T}^{\mu\nu}$ is the energy-momentum tensor. To obtain the wave equations, we suppose that the metric tensor in weak field can be written in the form:

$$\boldsymbol{g}_{\mu\nu} = \boldsymbol{\eta}_{\mu\nu} + \boldsymbol{h}_{\mu\nu} \quad (2)$$

[1] Greek indices run from 0 to 3, while Latin indices run from 1 to 3.

where $\boldsymbol{h}_{\mu\nu}$ is a small perturbation of the Minkowski tensor $\boldsymbol{\eta}_{\mu\nu}$ of flat spacetime ($\mathbf{h}^{\mu\nu} << \eta^{\mu\nu}$). By setting $\overline{\boldsymbol{h}}_{\mu\nu} = \mathbf{h}_{\mu\nu} - \frac{1}{2}\ \boldsymbol{\eta}_{\mu\nu}\mathbf{h}$, where $\boldsymbol{h} = \boldsymbol{h}^{\mu}_{\mu}$, Einstein's equation can be written in the form:

$$\Box\, \overline{\boldsymbol{h}}_{\mu\nu} = -\frac{16\pi G}{c^4}\boldsymbol{T}_{\mu\nu} \quad (3)$$

Where $\Box = \partial_\mu \partial^\mu = \nabla^2 - \frac{1}{c^2}\frac{\partial}{\partial t^2}$. In a vacuum ($\boldsymbol{T}_{\mu\nu} = \mathbf{0}$) the above equation becomes $\Box\, \overline{\boldsymbol{h}}_{\mu\nu} = \mathbf{0}$, whose solutions are gravitational waves propagating in empty space, which can be written in the form:

$$\overline{\boldsymbol{h}}_{\mu\nu} = \boldsymbol{A}_{\mu\nu} \cos(\boldsymbol{k}_\sigma \boldsymbol{x}^\sigma) \quad (4)$$

$$\boldsymbol{A}_{\mu\nu} = \boldsymbol{A}_{+}\begin{bmatrix} 0 & 0 & 0 & 0 \\ 0 & +1 & 0 & 0 \\ 0 & 0 & -1 & 0 \\ 0 & 0 & 0 & 0 \end{bmatrix} + \boldsymbol{A}_{\times}\begin{bmatrix} 0 & 0 & 0 & 0 \\ 0 & 0 & +1 & 0 \\ 0 & +1 & 0 & 0 \\ 0 & 0 & 0 & 0 \end{bmatrix} \quad (5)$$

Where $\boldsymbol{A}_{+}$ and $\boldsymbol{A}_{\times}$ are the amplitude of the wave in the two polarization states, and $\boldsymbol{k}^\sigma$is the four-plane wave vector $\boldsymbol{k}^\sigma = \left(\frac{\omega}{c}; \vec{\boldsymbol{k}}\right)$, where $\boldsymbol{\omega}$ is the frequency and $\boldsymbol{k} = \left|\left|\vec{\boldsymbol{k}}\right|\right| = \frac{\omega}{c}$ is the wave number, with $\boldsymbol{k}^\sigma \boldsymbol{k}_\sigma = \mathbf{0}$. This solution is given in the so-called TT gauge: the deformation caused by the wave is transverse to the progation direction, and the amplitude tensor $\boldsymbol{A}_{\mu\nu}$ is traceless. For instance, if we consider test masses positioned on a circle of radius R before the passage of the wave, the deformation provoked by the $\boldsymbol{A}_{+}$ polarization is given by:

$$\Delta \boldsymbol{S} = \boldsymbol{R}\left[\mathbf{1} + \frac{1}{2}\boldsymbol{A}_{+} \boldsymbol{cos}\,\omega \boldsymbol{t}\ \boldsymbol{cos}2\theta\right] \quad (6)$$

and it is depicted in Fig. 1, for a wave propagating along the z direction. A similar deformation is obtained for polarsiation $\boldsymbol{A}^{\times}$, which corresponds to a rotation of Fig1 by 45 degrees. Additionaly, the evolution of the deformation is depicted in Fig. 2.

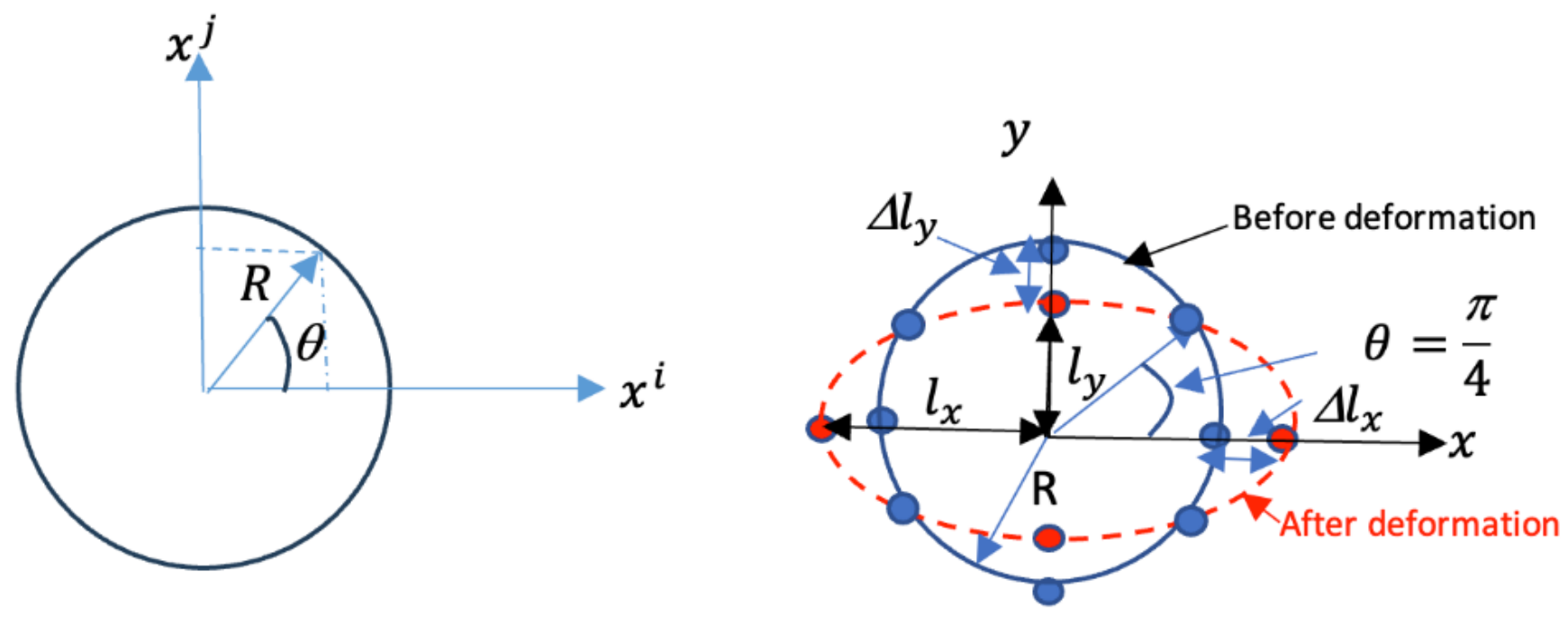

**Figure 1: Displacement of test masses in the plane $xy$ when a gravitational wave propagates in the direction $z$ –**

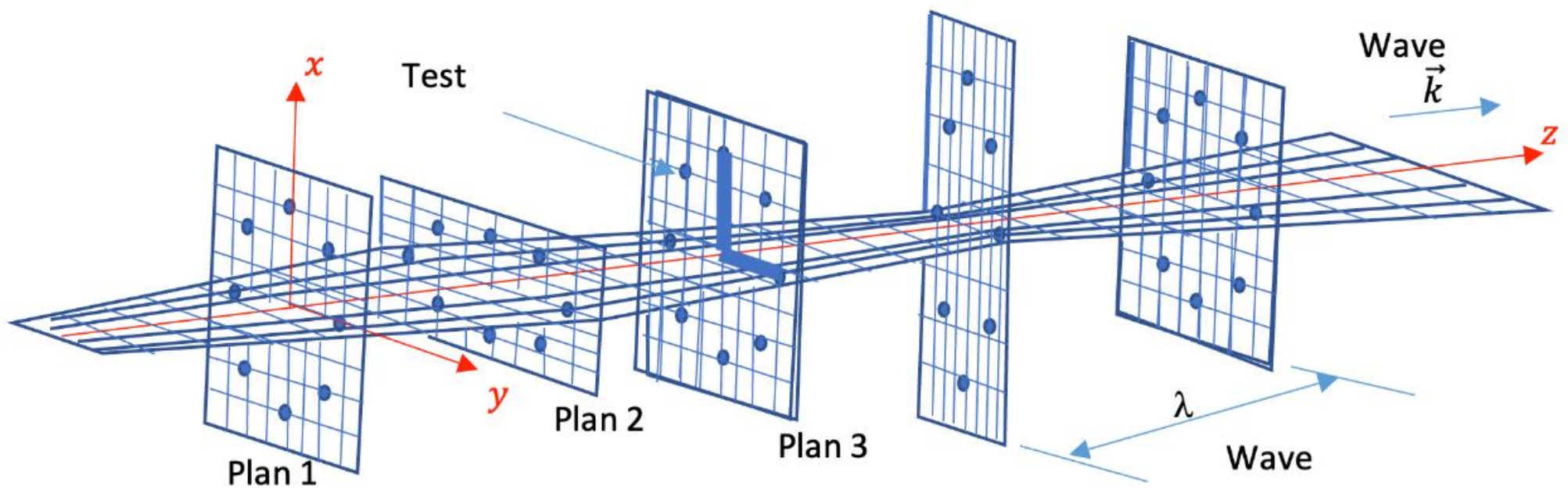

**Figure 2: Displacements of test masses when a gravitational wave passes in the direction $z$ followingsuccessive transverse planes –**

According to this description, the deformations provoked by the wave are in the plane orthogonal to the propagation direction: stricytly speaking, this is true if we confine ourselves to the first order with respect to the reference position. Up to second order (see e.g. D. Baskaran and L. P. Grishchuk **[34]** and M.L. Ruggiero in 2022 **[35]**), deformations also occur along the propagation direction. This effect (significantly enlarged) is depicted in Fig. 3.

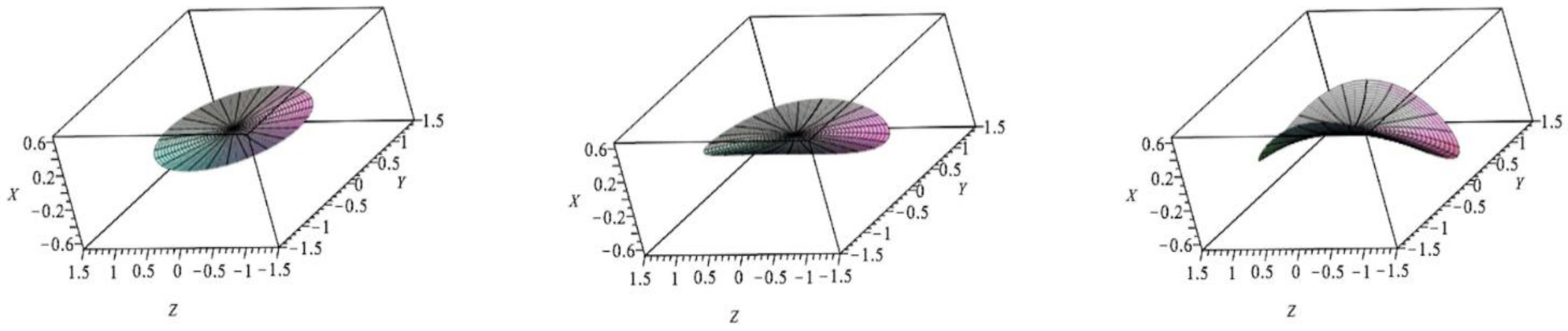

**Figure 3: Temporal evolution of test masses due to the effects of A+ polarization with up to second order [55] –**

### *3.2 Modified Theories of Gravity*

Altough General Relativity is the most successful model we have for understanding gravitational interactions, there are still unresolved issues regarding the application of GR to large-scale structures, such as the problems of dark matter and dark energy and. Additionaly, we do not yet know how to reconcile it with quantum mechanics. Consequently, several proposals have neen made to extend Einstein’s theory to address these issues. Many of these proposed theories have a richer geometrical structure. For instance, in Einstein-Cartan theory, torsion is present in addition to curvature. Specifically, torsion is related to the spin of the sources of the gravitational field **[56]**. For our purposes, it is interesting to note that this theory introduces additional polarizations for gravitational waves **[38]**. Our complementary polarizations appear as shown in Figure 4 and are associated with the following polarization matrices:

$$P_{\mu\nu}^{(+)} = \frac{1}{\sqrt{2}}\begin{pmatrix} 0 & 0 & 0 & 0 \\ 0 & 1 & 0 & 0 \\ 0 & 0 & -1 & 0 \\ 0 & 0 & 0 & 0 \end{pmatrix} \quad P_{\mu\nu}^{(\times)} = \frac{1}{\sqrt{2}}\begin{pmatrix} 0 & 0 & 0 & 0 \\ 0 & 0 & 1 & 0 \\ 0 & 1 & 0 & 0 \\ 0 & 0 & 0 & 0 \end{pmatrix} \quad (7)$$

$$P_{\mu\nu}^{(b)} = \frac{1}{\sqrt{2}}\begin{pmatrix} 0 & 0 & 0 & 0 \\ 0 & 1 & 0 & 0 \\ 0 & 0 & 1 & 0 \\ 0 & 0 & 0 & 0 \end{pmatrix} \quad P_{\mu\nu}^{(l)} = \frac{1}{\sqrt{2}}\begin{pmatrix} 0 & 0 & 0 & 0 \\ 0 & 0 & 0 & 0 \\ 0 & 0 & 0 & 0 \\ 0 & 0 & 0 & 1 \end{pmatrix} \quad (8)$$

$$P_{\mu\nu}^{(xz)} = \frac{1}{\sqrt{2}}\begin{pmatrix} 0 & 0 & 0 & 0 \\ 0 & 0 & 0 & 1 \\ 0 & 0 & 0 & 0 \\ 0 & 1 & 0 & 0 \end{pmatrix} \quad P_{\mu\nu}^{(yz)} = \frac{1}{\sqrt{2}}\begin{pmatrix} 0 & 0 & 0 & 0 \\ 0 & 0 & 0 & 0 \\ 0 & 0 & 0 & 1 \\ 0 & 0 & 1 & 0 \end{pmatrix} \quad (9)$$

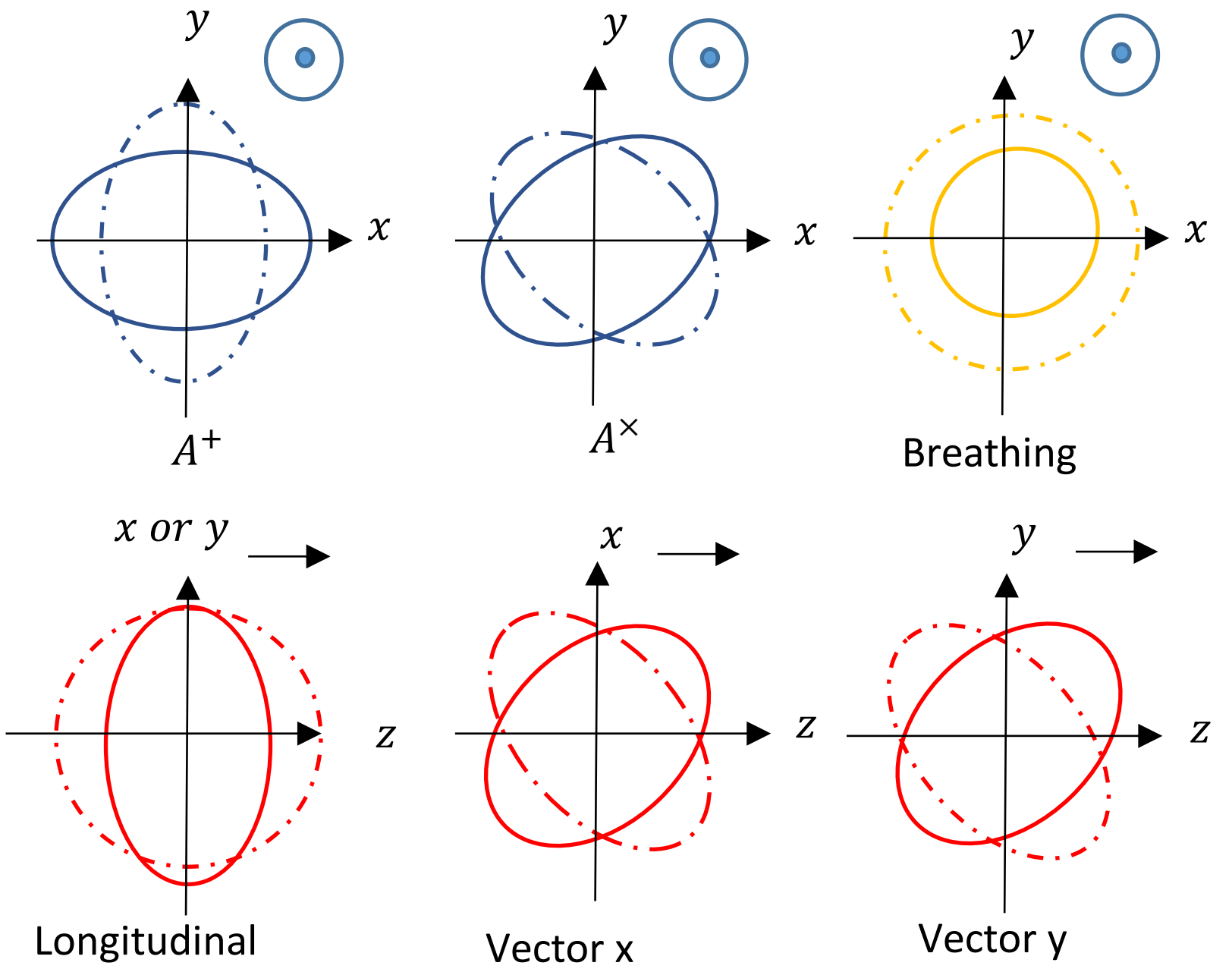

**Figure 4: Complementary polarizations that appear in the case of torsional modified general relativity in the case of the Einstein-Cartan-Sciamma-Kibble theory [38] –**

In particular, the first two matrices (45) are the classical ones predicted and measured by classical general relativity, the next four matrices (46) and (47), result from the introduction of the Einstein-Cartan-Sciamma-Kibble torsion.

More generally, as discussed by L. A. Philippoz **[39]**, and S. Mathur **[44]**, different alternative theories of gravity correspond to specific polarization features for gravitational waves, which are summarized in Table 1. Interestingly, all these modified theories converge on the same idea: that one or more polarizations complementary to $\boldsymbol{A}^{+}$ and $\boldsymbol{A}^{\times}$ should potentially exist. Only advanced measurements of deformations, by coupling various interferometers on Earth or using LISA, are likely to detect such polarizations. We will revisit this at the end of the paper.

| Theory | $+$ (*) | $\times$ (*) | $x$ (*) | $y$ | Breathing (*) | Longitudinal (*) |
|---|---|---|---|---|---|---|
| General relatvity | Yes | Yes | No | No | No | No |
| GR in noncompactified 4/6D Minkowski | Yes | Yes | Yes | Yes | Yes | Yes |
| Einstein/Aether | Yes | Yes | Yes | Yes | Yes | Yes |
| 5D Kaluza-Klein | Yes | Yes | Yes | Yes | Yes | No |
| Randall-Sundrum braneworls | Yes | Yes | No | No | No | No |
| Dvali-Gabadadze-Porrati braneword | Yes | Yes | Dep | Dep | Dep | Dep |
| Brans-Dicke Massive | Yes | Yes | No | No | Yes | Yes |
| Brans-Dicke Mass-less | Yes | Yes | No | No | Yes | No |
| F(R) Metric gravity | Yes | Yes | No | No | Yes | Yes |
| Bimetric Theory | Yes | Yes | Yes | Yes | Yes | Yes |
| Palatini Gravity | Yes | Yes | No | No | No | No |
| Scalar tensor theory | Yes | Yes | No | No | Yes | Yes |
| (*) See figure 4 for the definition of the type of polarisation $+ \times x\ y$ breathing and logitudinal | | | | | | |

**Table 1: Different polarizations associated with different theories of general relativity modified according to [39] [44] –**

Having completed the state of the art of gravitational wave polarizations in classical or modified general relativity, we will now see what the analogy of the continuum mechanics elasticity can help interpretating the results of general relativity (gravitational waves) recalled in the previous paragraphs.

### 3.3 Results of measurements made by LIGO/VIRGO interferometers

The first direct measurement of a gravitational wave occurred on September 14, 2015 (GW150914) and was presented on February 11, 2016 **[3]**. Two signals corresponding to the merger of two black holes that occurred 1.3 billion years ago were successively detected by the two LIGO interferometers in the USA. The nature of these signals is shown in Figure 5 below. The $x$-axis represents time in seconds, and the y-axis represents the h-deformations of space on the order of +/- $10^{-21}$.

The Figure 5 shows in red the almost perfect superposition of the two signals successively detected by the two interferometers 3000 km apart. Figure 7 shows the superposition of the theoretical curves and measured data after eliminating the noise. General relativity therefore predicts this type of signal remarkably well.

A second very interesting signal is GW170817 **[4]**, which corresponds to the merger of two neutron stars (kilonova) was observed simultaneously from both perspectives: gravitational waves and electromagnetic waves. This confirms, if needed, the remarkable efficiency and accuracy of the measurements. There can no longer be any doubt that rigid space deforms extremely little ($h_{xx} = -h_{yy} = 10^{-21}$) when stressed by large masses (black holes, neutron stars, etc.) concentrated in small volumes and rotating relative to each other at high speeds and high acceleration near the final coalescence time.

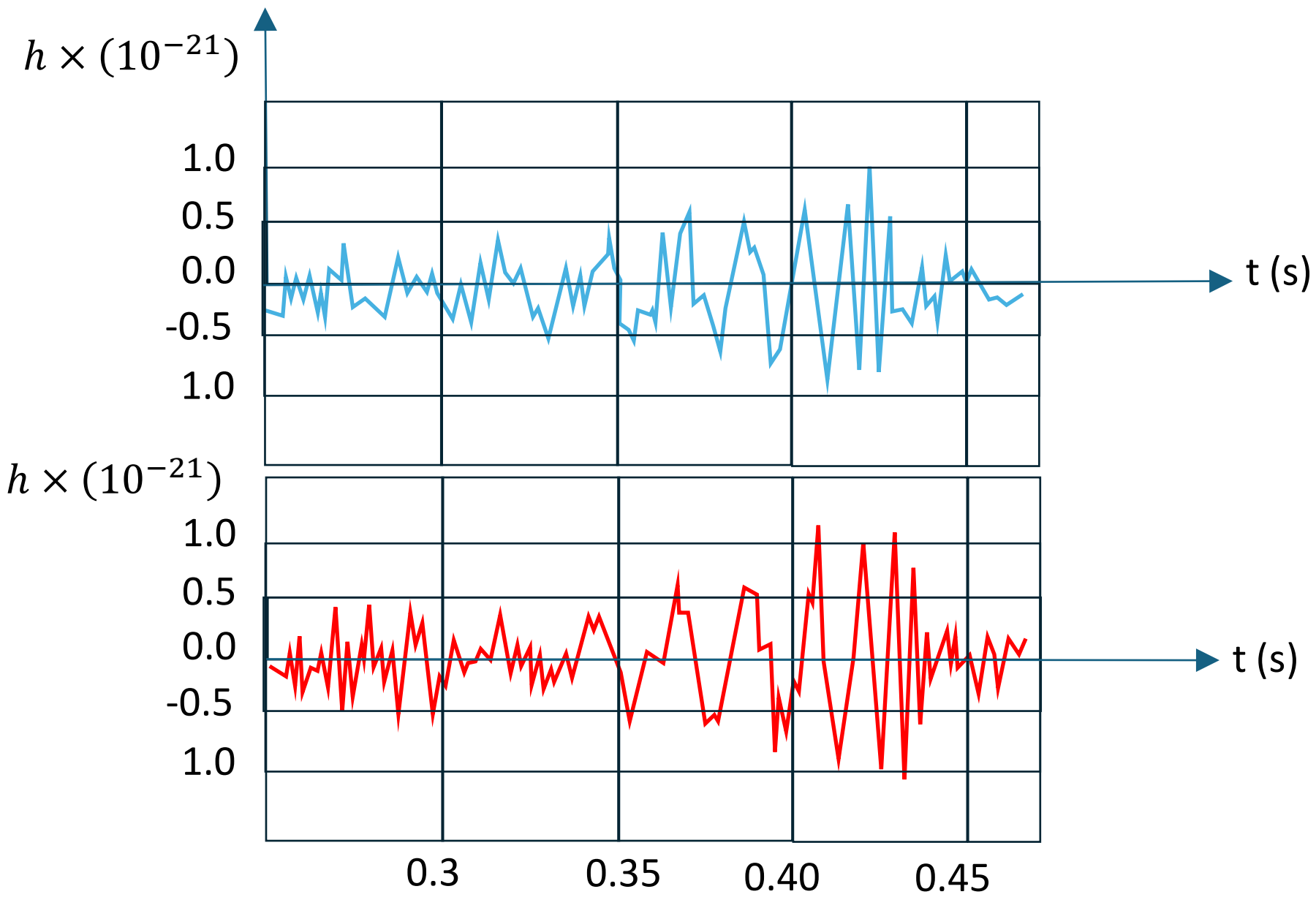

**Figure 5: Signal GW150914 picked up by the two LIGO laser interferometers sourced by B. P. Abbott et al. (LIGO Scientific Collaboration and Virgo Collaboration) – First Hanford second Livingstone [3] –**

*3.3 Discrepancies between LIGO/VIRGO interferometer measurements and the classical linearized theory of general relativity*

By carefully comparing the two GW150914 signals, it is possible to show the deviation in units $10^{-21}$ between the theory of general relativity and the interferometer measurements (Fig. 6). These curves demonstrate the precision of general relativity and indicate that any theory improving it should explain this few percent difference.

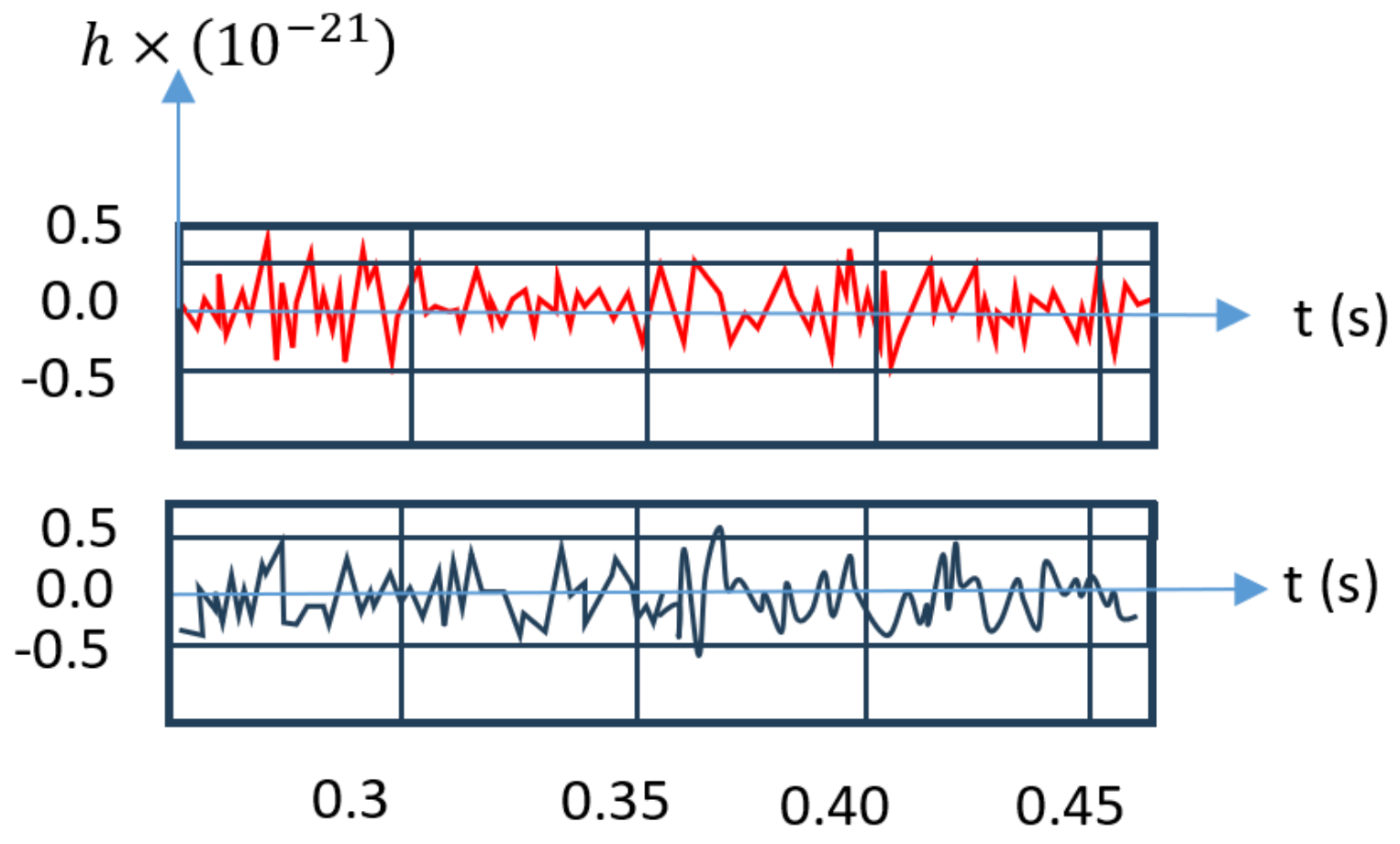

**Figure 6: Gap between the theory of general relativity and interferometer measurements for GW150914 source B. P. Abbott et al. (LIGO Scientific Collaboration and Virgo Collaboration) [3] –**

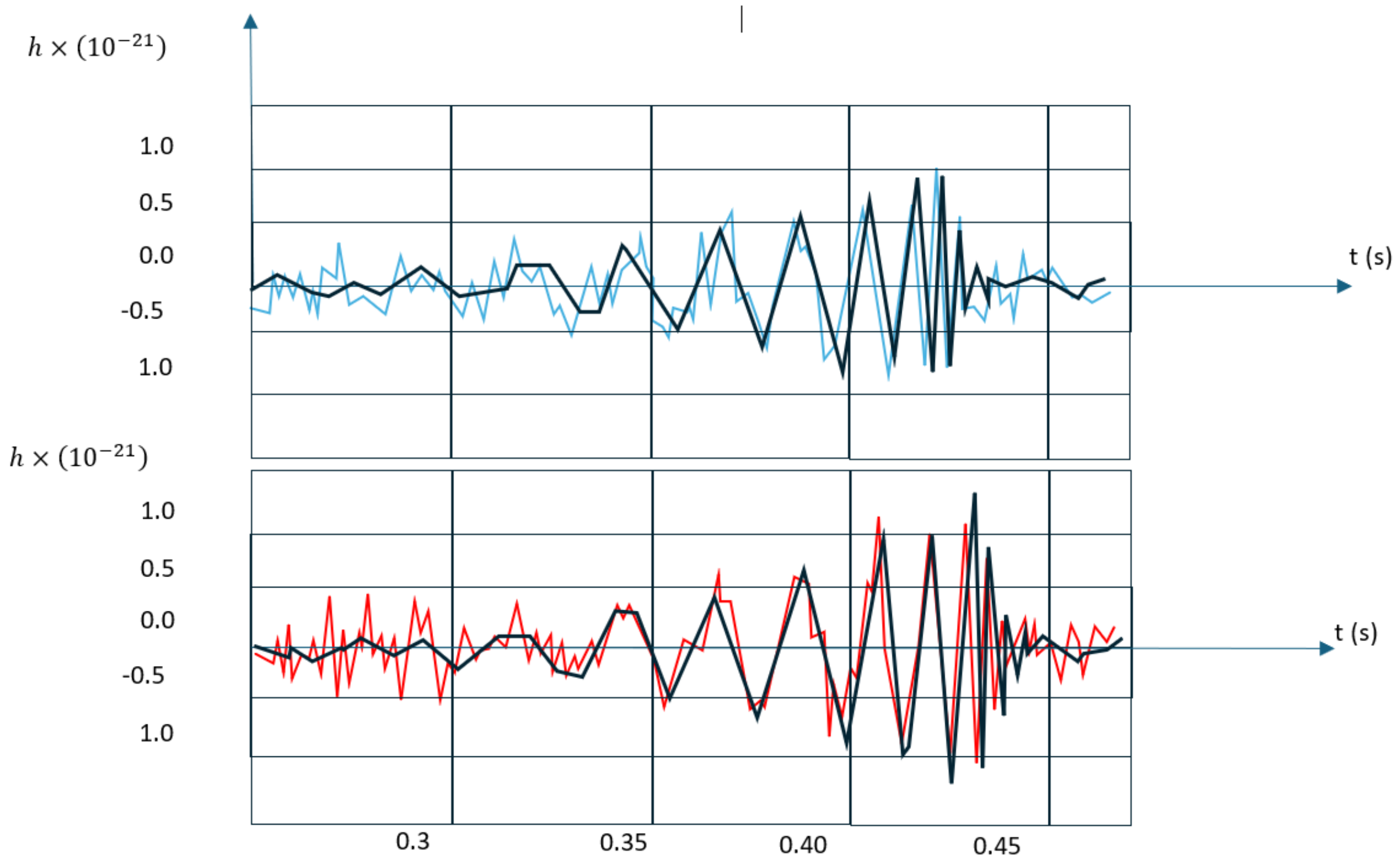

**Figure 7: Superposition of the theoretical curves (finite elements) and measured having eliminated the noise and interferometer measurements for GW150914 source B. P. Abbott et al. (LIGO Scientific Collaboration and VIRGO Collaboration) [3] -**

Viewing these results of extremely small deformations of space, physically measured by the two Ligo interferometers, which behave like giant strain gauges (each interferometer arm, traversed by a laser beam reflecting on various mirrors, is 4 km long and operates in a vacuum), it is conceivable to consider, by analogy, the latter as an elastic medium that deforms under the effect of energy or masses present within it.

Obviously, the continuum mechanics theory of elasticity seems appropriate for modeling these deformations of space by analogy.

In the following paragraphs, we will explore the strengths and limitations of this analogy, its potential contributions to our understanding of general relativity, and whether it can guide us in analyzing classical or modified theories of general relativity that aim to address the few percent difference between measurements and Einstein's version (Fig. 6 and Fig. 7).

## 4. Characterization of the deformations of space during the passage of a gravitational wave using different approaches of continuum mechanics

*4.1 Solid-state model for space*

*4.1.1 Case of stacking thin sheets in elasticity*

There are several important points to be made about this paper:

**a- Correspondence between polarizations established in general relativity and deformations established in the mechanics of continuous media**

The hypotheses, include, among other things, that in a weak field, the metric is expressed as :

$$\boldsymbol{g}^{\mu\nu} = \eta^{\mu\nu} + \boldsymbol{2\varepsilon}^{\mu\nu} \quad (10)$$

Where $\boldsymbol{\varepsilon}^{\mu\nu}$ is the equivalent of a first gradient strain tensor comprising spatial and temporal strain components (00, 01, 02, 03).

Comparing expression (10) with expression (2), the following equivalence emerges:

$$\boldsymbol{2\varepsilon}^{\mu\nu}\ (\boldsymbol{in\ mechanics})\ = \boldsymbol{h}^{\mu\nu}\ (\boldsymbol{in\ GR\ in\ weak\ field})\ (11)$$

This correspondence is fundamental for the rest of this paper, as it implies that the components of the polarization matrices of gravitational waves can be interpreted, within the context of the analogy with the elastic medium, as the components of a deformation tensor of the associated elastic cosmic space **[9]**, **[11]**, **[12]**.

**b- Relations between longitudinal and transversal strains**

Since gravitational waves are transverse, (no longitudinal wave), and the deformations are of equal intensity but opposite sign in each arm of the interferometers, this implies Poisson's ratios of the equivalent anisotropic elastic cosmic medium as:

$$\nu_{xy} = \nu_{yx} = 1\ (12)$$

There are two ways to obtain this value. First, measurements are made on the LIGO/VIRGO interferometers, where elongation and shortening are measured simultaneously in each of the arms with the same intensity and opposite signs (Fig. 5). Alternatively, we can assume an elastic medium and impose the absence of a longitudinal wave (since gravitational waves are transverse according to classical general relativity). Indeed, the velocity of a compression wave in an anisotropic elastic medium can be written as:

$$c_{pressure} = \sqrt{\frac{Y_{xy}(1-\nu_{xy})}{\rho(1+\nu_{xy})(1-2\,\nu_{xy})}} = 0,\ then\ \nu_{xy} = 1 \quad (13)$$

For the other variation $\nu_{yx} = 1$, that is observed in the strains of each arms (Figures 1 and 5).

We therefore find that, as interferometers are in the plane $xy$: $\nu_{xy} = \nu_{yx} = 1$

This observation regarding the values of these Poisson's ratios implies that:

- The distortions of space caused in the planes $x, y$ perpendicular to the direction $z$ of propagation of the gravitational waves (Fig. 2), associated with polarizations according to the principle described above, are of the same magnitude but of the opposite sign,
- On one hand, the elastic medium associated with space in continuum mechanics is necessarily anisotropic due to the value of these Poisson's ratios. On the other hand, we note the absence of polarisation and, thus, complementary deformation in the z direction of wave propagation in classical general relativity,
- According to this result, this equivalent elastic spatial medium would consist of stacking of sheets that deform independently of each other ($\nu_{xz} = \nu_{yz} = 0$). The spatial medium thus consists more in a tailored medium at a very small scale, than an equivalent homogeneous medium (Fig. 2), **[60]** where we have strains in three directions of space, not only in two directions $(x, y)$ as in classical general relativity.

Note that in the case of a gravitational wave not perpendicular to the interferometer arms, the angle $\theta$ of the observer is taken into account as follows **[58]**:

$$h_{+(t)} = \frac{4G\mu a^2 \omega^2}{Rc^4}\left(\frac{1+cos^2\theta}{2}\right)\cos(2\omega t)(14a)$$

$$h_{\times(t)} = \frac{4G\mu a^2 \omega^2}{Rc^4} cos\theta \sin(2\omega t)(14b)$$

With: $\omega = \sqrt{\frac{GM}{a^3}}$ ; $M = m_1 + m_2$ ; $\mu = \frac{m_1 m_2}{M}$

θ: the angle between the normal to the rotation plane of the two bodies and the direction R of the observer,

a: the distance between the two rotating bodies (eg black holes),

*R*: the distance between the observer and the system in rotation (*R* >> a),

c: the speed of light and G: the gravitational constant.

**c- The Young's moduli of the equivalent elastic medium can be expressed in terms of fundamental constants**

This model was developed by T. G. Tenev, and M. F. Horstemeyer **[9]**, among others. In this paper, cosmic space is assumed to consist of ultra-thin sheets whose thickness is that of the Planck length, denoted $l_p$.

The consequence is that T. G. Tenev, and M. F. Horstemeyer in **[9]** arrive at these expressions of Young's moduli of cosmic fabric:

$$\boldsymbol{Y = Y_x = Y_y = E_x = E_y = \frac{6c^7}{2\pi\hbar G^2} = \frac{24}{l_p^2 \kappa}} \quad (15)$$

$Y_z$ not being defined in their model.

With $\hbar$ Planck's reduced constant, G is the gravitational constant $l_p$ is the Planck length, and c is the speed of light.

The cosmic fabric in this paper is an extension of the cosmic medium as seen by **[9]** with a transversal isotropic behavior.

The numerical application leads to values of the Young's moduli $Y_x$ and $Y_y$ of the cosmic fabric, (in correlation with the energy of the vacuum according to Sackarov **[5]**) that are outside the usual standards if we assume a stacking of space sheets of the Planck thickness constituting this cosmic fabric.

$$Y_{x(Vacuum)} = Y_{y(Vacuum)} = 4.4 \times 10^{113} Pa$$

This value is associated with a vacuum density of $\rho_{vacuum}$ =1.3×$10^{96}$ kg/$m^3$, which is so extremely and lacks a clear understanding.

Two additional remarks emerge from this analysis:

The black hole radiation equation of Hawking is one of the few relations that combine the three fundamental constants c, h, and G in addition of the Boltzman constant $k_B$. This can be achieved through the analogy of the equivalent elastic medium via the Y Young's modulus with equation (15). Thus, the elastic analogy provides a way to combinate these three fundamental physic constants.

By reversing relation (15), the gravitational constant G becomes dependent on the elastic characteristics of the equivalent cosmic elastic medium. In their approaches, T.G. Tenev and M.F. Horstemeyer did not consider an anisotropic medium with several Young's moduli. Thus they propose:

$$G = \sqrt{\frac{6c^7}{hY_k}} \quad (16)$$

The elastic medium associated with space in continuum mechanics is necessarily anisotropic due to the values of these Poisson's ratios (transverse isotropic in the plan $x, y$). It is also anisotropic due to the absence of $z$-polarization, which is connected with the deformation of space as seen on equation (11). $Y_k$ is the Young's modulus in the direction k.

### *4.1.2 Case of the deformation study of two 90° space tubes containing the interferometer arms*

D. Izabel in **[11]** and **[12]** generalized the work of T. G. Tenev, and M.F. Horstemeyer in **[9]** (Fig. 8) regarding a medium consisting of several thin sheets of Planck thickness each characterized by an associated Young's modulus $Y = Y_x = Y_y$ and energy. He sought the analogy with the mechanics of continuous media not by modifying the field equation of general relativity but by introducing mechanical parameters into the constant of proportionality κ, which becomes the flexibility characteristics of the cosmic fabric expressed as a function of Young's modulus. He finds there a formulation analogous to that of a generalized Hooke's law.

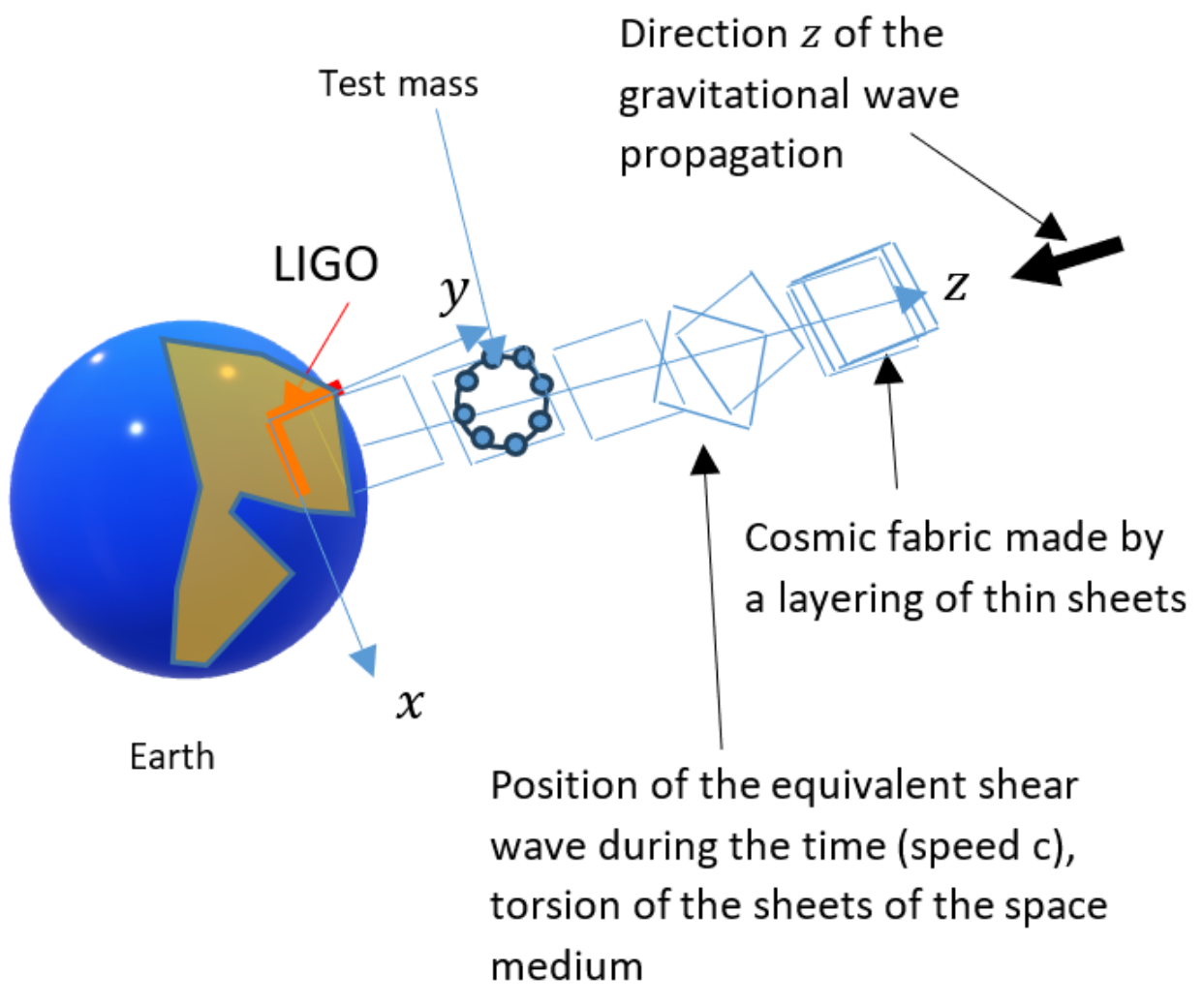

**Figure 8: Visualization of the successive deformations of a stack of space sheets during the passage of a gravitational wave**

–

Thus, in the case of intermeters positioned in a plane of this cosmic sheet are, in each arm he is able to **[11]**:

$$\frac{1}{L^2}(\varepsilon_{xx})^2 = 4(1 + \nu_{xy})\pi \frac{\pi f^2}{\rho} \frac{1}{c^4} \frac{U}{V} \, (17a)$$

$$\frac{\mathbf{1}}{\boldsymbol{L}^{\mathbf{2}}}\left(\boldsymbol{\varepsilon}_{\boldsymbol{yy}}\right)^{2}=\mathbf{4}\left(\mathbf{1}+\boldsymbol{\nu}_{\boldsymbol{yx}}\right)\boldsymbol{\pi}\frac{\boldsymbol{\pi}\boldsymbol{f}^{\mathbf{2}}}{\boldsymbol{\rho}}\frac{\mathbf{1}}{\boldsymbol{c}^{\mathbf{4}}}\frac{\boldsymbol{U}}{\boldsymbol{V}}\quad(\mathbf{17b})$$

These are a mechanical version of the Einstein's field equation (1) in space.

L is the length of the interferometer arm, U is the elastic deformation energy of the volume V of the arm, f is the natural frequency of vibration of the space in the arm in accordance with the gravitational wave, ρ is the density of the vacuum, $\varepsilon_{xx}$ and $\varepsilon_{yy}$ are the deformations of the space in the arms, with $v = \nu_{xy} = \nu_{yx} = 1$, the Poisson's ratios in the plane $xy$.

Assuming equivalence in curvature tensor (the radius of curvature R locally tends to $1/L^2$ so, is quasi flat as for the curvature k of the universe) we can write in the plane of the interferometers:

$$R_{ij}=\begin{bmatrix}\frac{1}{L^2} & 0\\ 0 & \frac{1}{L^2}\end{bmatrix}\begin{bmatrix}(\varepsilon_{xx})^2 & 0\\ 0 & \left(\varepsilon_{yy}\right)^2\end{bmatrix}\quad(18)$$

And defining $\frac{U}{V} = T_{xx} = T_{yy}$ the components of the strain energy tensor of the cosmic fabric, with U as the strain energy and V as the volume of the interferometer arms.

$$T_{ij}=\begin{bmatrix}T_{xx} & 0\\ 0 & T_{yy}\end{bmatrix}(19)$$

If we use $v = \nu_{xy} = \nu_{yx} = 1$ , as explained in 5.1.1, in the expression (20) below:

$$\mathbf{4}(\mathbf{1}+\boldsymbol{\nu})\,\boldsymbol{\pi}\,\frac{\boldsymbol{\pi f^2}}{\boldsymbol{\rho}}\,\frac{\mathbf{1}}{\boldsymbol{c^4}}=\mathbf{8}\,\boldsymbol{\pi}\frac{\boldsymbol{G}}{\boldsymbol{c^4}}\quad(\mathbf{20a})$$

$$\boldsymbol{G}=\frac{\boldsymbol{\pi f^2}}{\boldsymbol{\rho}}=\frac{\boldsymbol{\pi f^2 c^2}}{\boldsymbol{Y}}\quad(\mathbf{20b})$$

We obtain Einstein's expression (21) in weak field general relativity transposed into an equivalent elastic medium in the plane of the interferometers as:

$$\boldsymbol{R_{ij}}=\frac{\boldsymbol{8\pi G}}{\boldsymbol{c^4}}\boldsymbol{T_{ij}}=\boldsymbol{\kappa T_{ij}}\quad(\mathbf{21})$$

**Remark**

It can be shown **[11]**, **[12]** that the expression (20a) for the cosmic fabric flexibility κ is-related to this expression of G as a function of the squared frequency of vibration of the space in the tube and the energy density of the vacuum in quantum field theory.

Expression (20b) developped in **[11]** and **[12]** shows that the Young's modulus Y can be expressed as a fonction of G, f and c.

The conclusion of this study is that the expression of Einstein's general relativity can be seen in planes $xy$ transverse to the direction of propagation of the gravitational wave as a Hooke's law, with κ playing the role of the flexibility of the equivalent cosmic fabric structure, but with an anisotropy that remains non-standard.

#### *4.1.3 Case of a torsionally stressed space cylinder*

In publications **[11]** and **[12],** the author uses the analogy between the perturbations of the metric and the associated distortions of space. Thus, the 2 classical polarizations $A^{+}$ and $A^{\times}$ of gravitational waves can be read as follows:

$$h_{\mu\nu} = A_{+}\cos\left(\frac{\omega}{c}(ct-z)\right)\begin{bmatrix}0 & 0 & 0 & 0\\ 0 & +1 & 0 & 0\\ 0 & 0 & -1 & 0\\ 0 & 0 & 0 & 0\end{bmatrix} \rightarrow \varepsilon_{xy\,(A_{+})} = \frac{1}{2}A_{+}\cos\left(\frac{\omega}{c}(ct-z)\right)\begin{bmatrix}\varepsilon_{xx} & 0 & 0\\ 0 & -\varepsilon_{yy} & 0\\ 0 & 0 & 0\end{bmatrix} \quad (22)$$

These two components of deformations correspond to elongations and shortenings, and:

$$h_{\mu\nu} = A_{\times}\cos\left(\frac{\omega}{c}(ct-z)\right)\begin{bmatrix}0 & 0 & 0 & 0\\ 0 & 0 & +1 & 0\\ 0 & +1 & 0 & 0\\ 0 & 0 & 0 & 0\end{bmatrix} \rightarrow \varepsilon_{xy\,(A_{\times})} = \frac{1}{2}A_{\times}\cos\left(\frac{\omega}{c}(ct-z)\right)\begin{bmatrix}0 & \varepsilon_{xy} & 0\\ \varepsilon_{yx} & 0 & 0\\ 0 & 0 & 0\end{bmatrix} \quad (23)$$

These two components of deformations correspond to distortions, which in elasticity correspond to pure torsion. (Fig. 9).

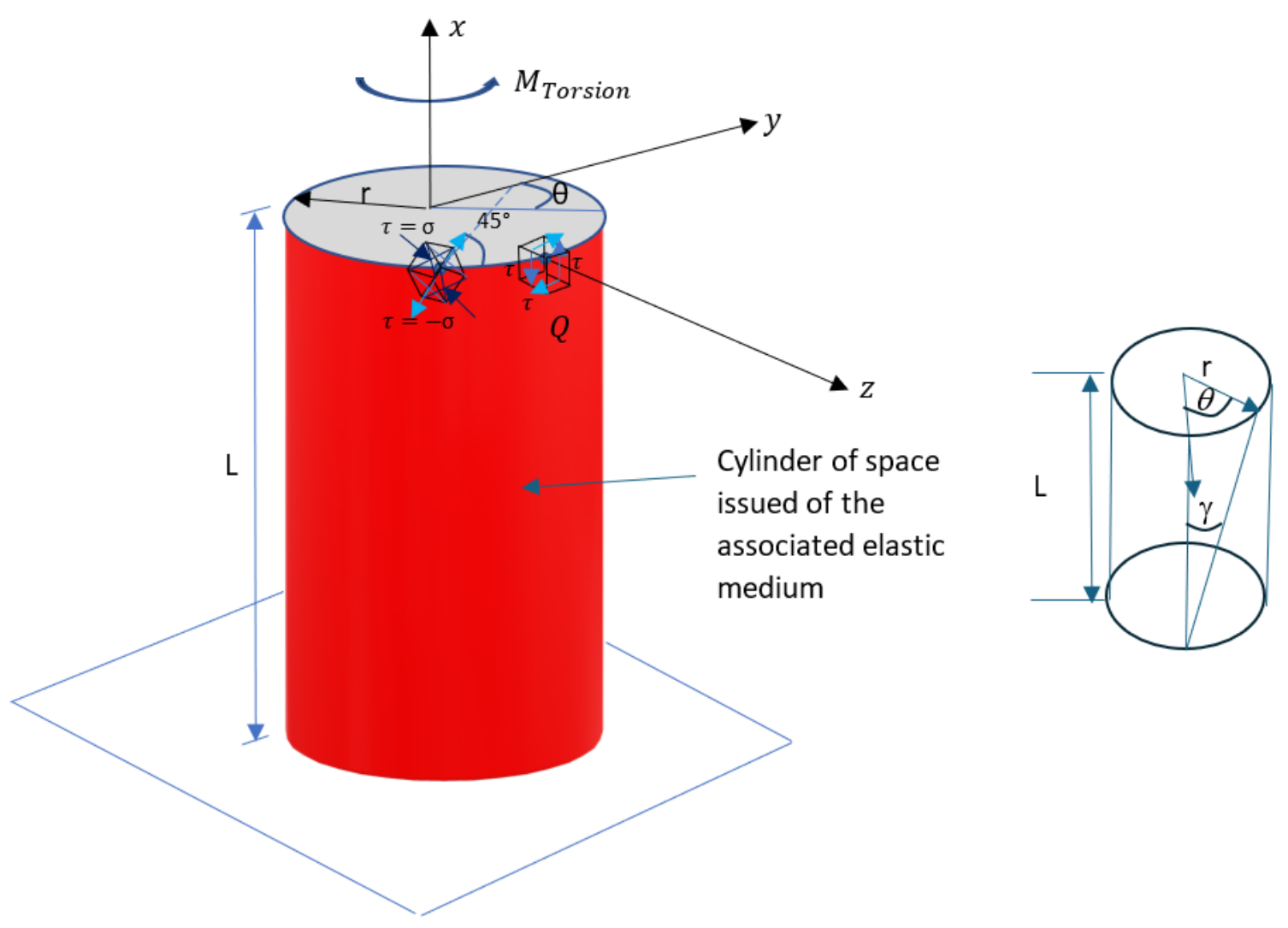

**Figure 9: Pure torsion state of an isotropic material in elasticity theory –**

Gravitational waves are caused by the rotation of binaries (two black holes, two neutron stars, one black hole, one neutron star, etc.) that can be seen as a twist of space due to their rotation relative to each other (Fig. 10).

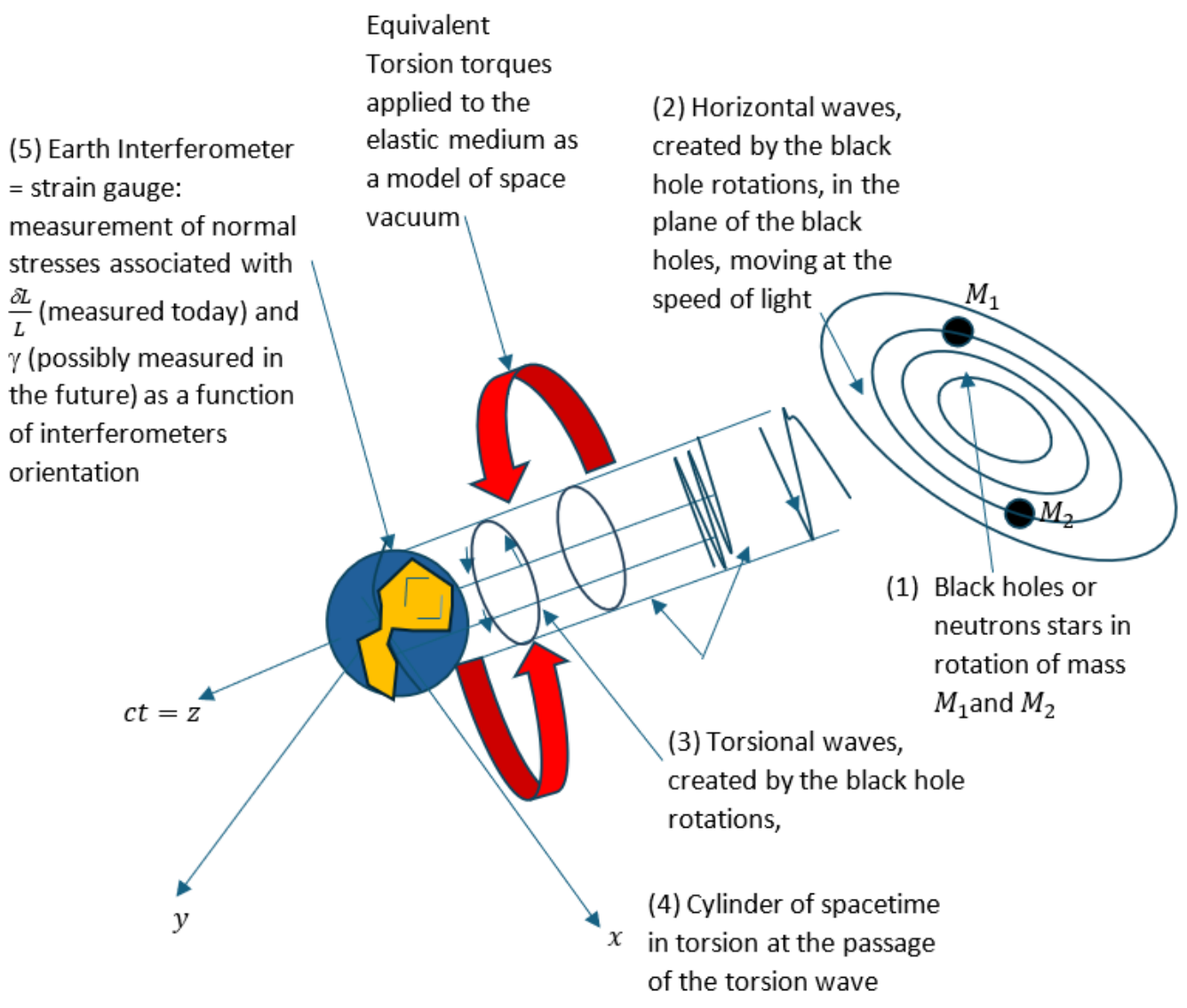

**Figure 10: Effect of the rotation of a binary following the analogy of the torsional elastic medium –**

However, if we use the analogy in the other direction by asking the following question: can an elastic reading of the two polarizations obtained in general relativity offer us additional information in the weak field? By examining the two deformation tensors above, we observe elongations $\varepsilon_{xx}$ and $\varepsilon_{yy}$ shortenings in the plane $xy$ associated with the deformations and, already widely measured with the LIGO/VIRGO interferometers. Depending on the elasticity, angular distortions in 45° planes would also be possible associated with the deformations $\varepsilon_{xy}$ and $\varepsilon_{yx}$ (Fig. 11) and **[11]**. It is not possible to measure these distortions by current interferometers because they are not designed for this. This should be possible using the LISA interferometer or the future 3-arm triangle Einstein telescope or multiple pulsars as was done for the 2023 detection of the stochastic gravitational wave background.

$$tan\,\gamma_{ij} \approx \gamma_{ij} = \frac{b}{L} (24)$$

Finally, in **[11]** by studying a cylinder of elastic space twisted, the author finds a mechanical version of general relativity in which it again appears $G = \frac{\pi f^2}{\rho}$. The deformations associated with the curvature are angular distortions, noted as γ in the Figure 11, where T is the strain energy U divided by the volume V of the interfermoters arms.

$$U = \frac{1}{2}\int_0^L \frac{M_t^2}{\mu I_t} dx \quad (25)\ ;\ \frac{\mathbf{1}}{\boldsymbol{L^2}}\boldsymbol{\gamma^2} = \mathbf{16}\boldsymbol{\pi}\frac{\boldsymbol{G}}{\boldsymbol{c^4}} \times \boldsymbol{T} \quad (26)$$

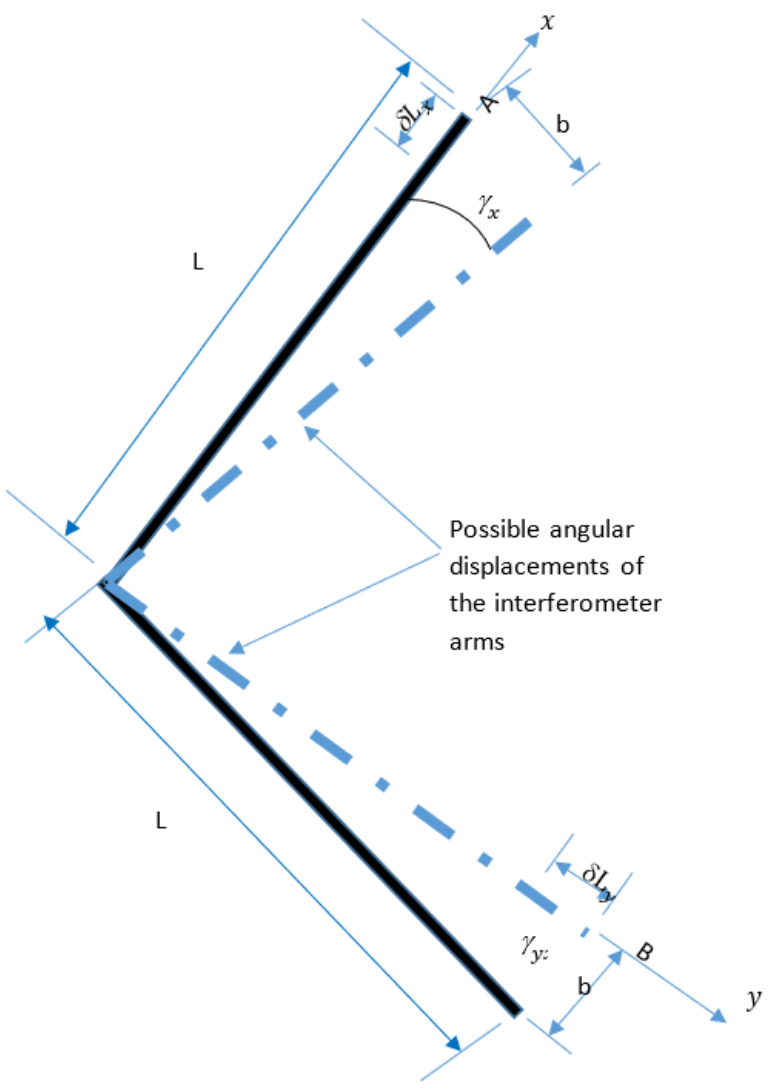

**Figure 11: Illustration of a non-measurable angular distortions of the arms interferometers in the plane $xy$ (top view) –**

This can be compared with (3) by replacing $h_{\mu\nu}$ with (11) using this mechanical expression of general relativity in the weak field:

$$\square\left(2\varepsilon_{\mu\nu}+\frac{1}{2}\eta_{\mu\nu}2\overline{\varepsilon}\right)=-\frac{16\pi G}{c^4}T_{\mu\nu} \quad (27)$$

## 5. Analysis and discussion of the results obtained according to general relativity and the analogy of the elastic medium

### *5.1. Concerning the analogy between the polarizations of gravitational waves and the different forms of the associated deformation tensor of space*

One of the interesting contributions of the elastic medium analogy is the opportunity to interpret, according to **[9]**, **[11]** and **[12]**, the polarizations of gravitational waves in linearized general relativity as components of a space deformation tensor by applying the analogy of the elastic medium to cosmic space.

This provides a new illustration of the two polarizations $A^{+}$ and $A^{\times}$. In the analogy of the elastic medium sujected to ~~in~~ pure torsion by two massive rotating objects (e.g. GW150914), the strain tensor consists of four components. two diagonal components corresponding to facets stressed in tensile and compressive motion, and two components in the other corresponding diagonal for another facet at 45° of the previous normal stress (Fig. 12). Due to the Mohr circle associated with this pure torsion, these shear stresses τ are of the same intensity as the normal stresses σ on the other facet.

### *5.2. Possible lateral deformations of the interferometers*

According to **[11]**, these potential angular deformations could be easily calculated or measured geometrically using the Mohr circle ~~to~~ for two of the arms of the future LISA experiment, as shown in Figure 12, **[59]**:

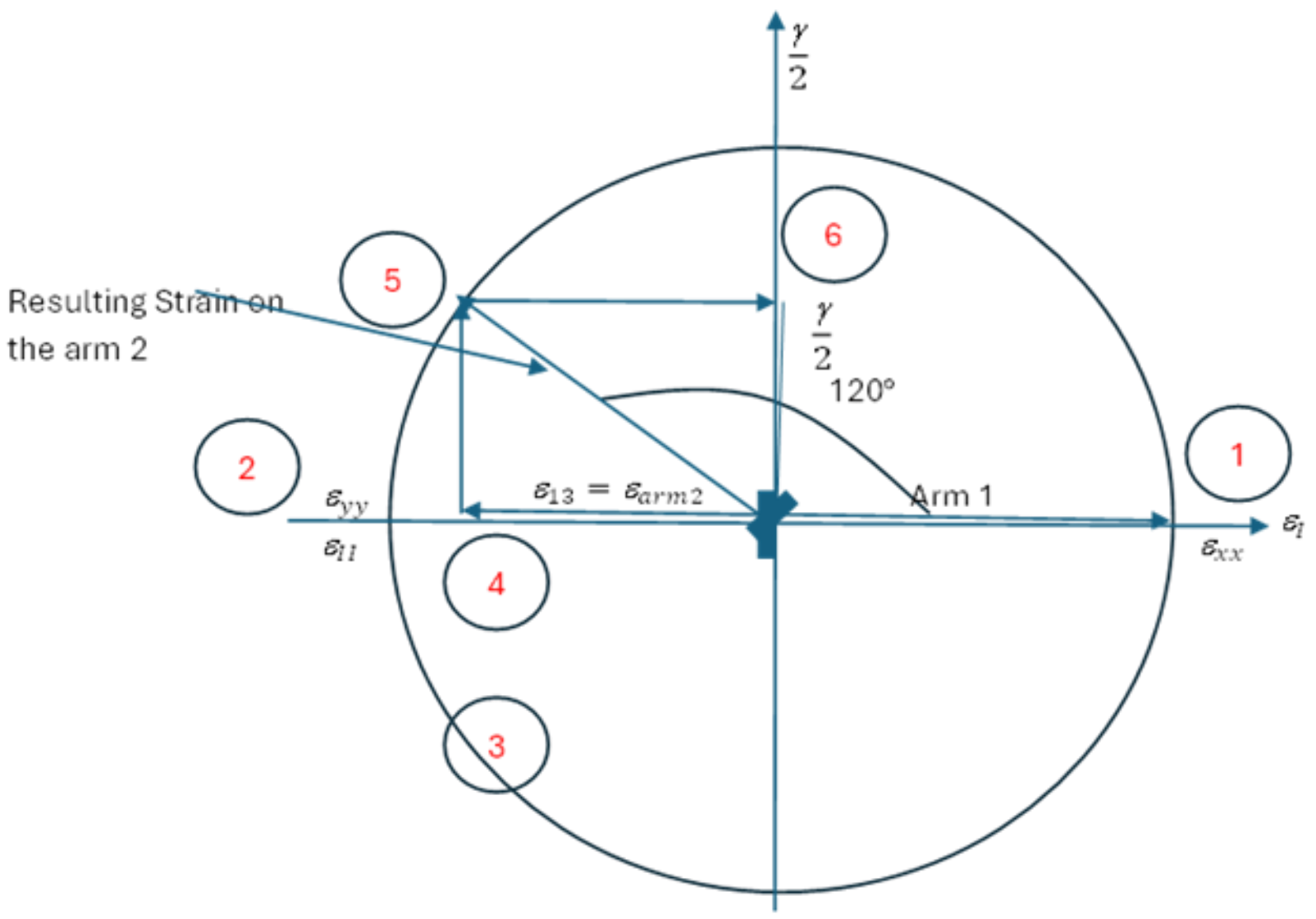

**Figure 12: Determination of a potential angular distortion from the future LISA (arms at 60°) –**

The layout is as follows (Fig. 12):

1. Plot the deformation $\varepsilon_{xx}$ of Arm 1 from the measurement of the elongations of this arm, calculated from the time it takes for the laser beam to travel the distance between Satellite 1 and Satellite 2.
2. Trace the deformation $\varepsilon_{yy} = -\varepsilon_{xx}$,
3. Plot the Mohr circle passing through $\varepsilon_{xx}$ and $\varepsilon_{yy}$,
4. The deformation $\varepsilon_{\text{Arm 2}}$ of Arm 2 is reported from the measurement of the shortening of this arm (this is a shortening slightly less than the maximum shortening located at 90° (stated by Ligo and Virgo),
5. Draw the direction of Arm 2 by rotating -120° (due to the angle of 60° between the arms of the future LISA interferometer) the direction of the vector associated with the facet of Arm 2,
6. From this, we derive the second component (half distortion $\gamma/2$) for the facet associated with Arm 2 on the vertical axis.

The value of the distortion is intersected by the lengthening and shortening of Arm 3 with the expression α seen between the arms below (Fig. 13).

$$\cos\alpha_{(t)} = \frac{{L_2}^2 + {L_1}^2 - L_3^2}{2L_1L_2} = 60° + \gamma' \quad (28)$$

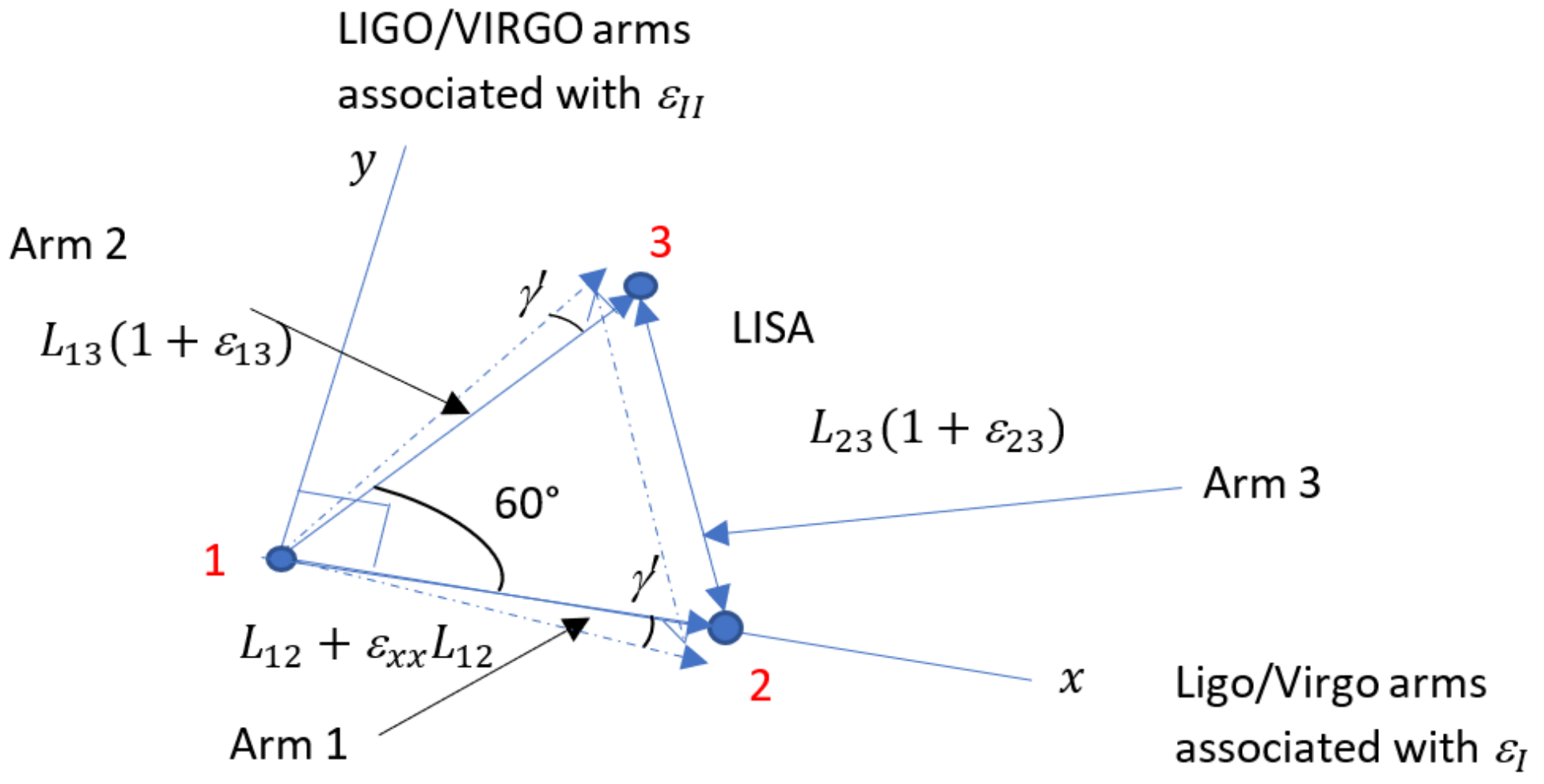

**Figure 13: Determination of angles α and γ' from variations in LISA arms length measurements –**

### *5.3. Concerning the anisotropy of space based solely on current general relativity*

We have seen that linearized general relativity implies, in the case of gravitational waves, two unique polarizations that can be read as two expressions of a strain tensor involving a Poisson's ratio of 1 in these distorted transverse planes **[11]** and **[9]**. Since general relativity does not predict a longitudinal wave, the associated space model is found in plane deformations without any deformations in the direction of wave propagation involving zero Poisson's ratios in all planes including the direction of propagation. In this analogy, the elastic medium seems to be made up of transverse plane independent of each others. From a mechanical point of view, it seems clear that, if we adhere to general relativity in its current version, that this blatant anisotropy is contradictory to the hypothesis generally made in physics of a homogeneous and isotropic transverse medium. Clearly, in our analogy, 3-dimensional elastic space can no longer be isotropic. We are missing the deformations $\varepsilon_{xz}$, $\varepsilon_{yz}$.

We therefore have, in the case of this local transverse isotropy and the notations of the ASTER code:

For the Young's moduli:

$$Y_L = E_L = Y_T = E_T \qquad (29)$$

For shear moduli:

$$G_{TN} = G_{LN} \qquad (30)$$

$$G_{LT} = \frac{E_L}{2(1+\nu_{LT})} \quad (31)$$

On the basis of deformations in the planes perpendicular to the direction of propagation of gravitational waves in weak field general relativity, unmodified:

$$\frac{\nu_{LN}}{E_L} = \frac{\nu_{TN}}{E_L} = 0 \quad (32)$$

Therefore:

$$\nu_{NT} = \nu_{NL} = 0 \quad (33)$$

$$\nu_{LN} = \nu_{TN} = 0 \quad (34)$$

While in the plane $xy$:

$$\nu_{LT} = \nu_{TL} = 1 \quad (35)$$

From the displacements and associated strains imposed at the cosmic fabric by the mass or energy present in space or for vacuum cosmic fabric subjected to the gravitational wave elastic energy **[40]**, **[41]**, it is thus possible to define the equivalent stress field (see introduction) as described in formulas (36), (37), and (38).

Thus, writing the generalized Hooke's law in the frame of reference (L, T, N), where N is he direction of propagation as $\hat{\varepsilon} = K^{-1}\hat{\sigma}$, with:

$\hat{\varepsilon}^T = (\varepsilon_{LL}, \varepsilon_{TT}, \varepsilon_{NN}, 2\varepsilon_{LT}, 2\varepsilon_{LN}, 2\varepsilon_{TN})$ and $\hat{\sigma}^T = (\sigma_{LL}, \sigma_{TT}, \sigma_{NN}, \sigma_{LT}, \sigma_{LN}, \sigma_{TN})$, $K^{-1}$ is the compliance matrix. While the definition of the strain tensor is clear with the space deformation theory, the same cannot be said for the stress tensor $\hat{\sigma}^T$. In our case, we define this stress tensor as the equivalent stress field that induces the observed strain tensor in linearized elasticity under small strain.

The generalized transverse isotropic Hooke's law is:

$$\begin{Bmatrix} \varepsilon_{LL} \\ \varepsilon_{TT} \\ \varepsilon_{NN} \\ 2\varepsilon_{LT} \\ 2\varepsilon_{LN} \\ 2\varepsilon_{TN} \end{Bmatrix} = [K^{-1}] \begin{Bmatrix} \sigma_{LL} \\ \sigma_{TT} \\ \sigma_{NN} \\ \sigma_{LT} \\ \sigma_{LN} \\ \sigma_{TN} \end{Bmatrix} \quad (36)$$

With $[K^{-1}]$, the classical compliance matrix at the point M in the frame of reference (L, T, N*):* M (L, T, N)) *(37)*

$$\begin{bmatrix} \frac{1}{E_L} & \frac{-\nu_{LT}}{E_L} & \frac{-\nu_{LN}}{E_L} & 0 & 0 & 0 \\ \frac{-\nu_{TL}}{E_T} & \frac{1}{E_T} & \frac{-\nu_{TN}}{E_T} & 0 & 0 & 0 \\ \frac{-\nu_{NL}}{E_N} & \frac{-\nu_{NT}}{E_N} & \frac{1}{E_N} & 0 & 0 & 0 \\ 0 & 0 & 0 & \frac{2(1+\nu_{LT})}{E_L} & 0 & 0 \\ 0 & 0 & 0 & 0 & \frac{1}{G_{LN}} & 0 \\ 0 & 0 & 0 & 0 & 0 & \frac{1}{G_{TN}} \end{bmatrix} M\ (L,T,N))$$

In the case of classical gravitational waves, $K^{-1}$ becomes (38):

$$\begin{bmatrix} \frac{1}{E_L} & \frac{-1}{E_L} & 0 & 0 & 0 & 0 \\ \frac{-1}{E_L} & \frac{1}{E_L} & 0 & 0 & 0 & 0 \\ 0 & 0 & \frac{1}{E_N} & 0 & 0 & 0 \\ 0 & 0 & 0 & \frac{4}{E_L} & 0 & 0 \\ 0 & 0 & 0 & 0 & \frac{1}{G_{LN}} & 0 \\ 0 & 0 & 0 & 0 & 0 & \frac{1}{G_{LN}} \end{bmatrix} M\ (N,L,T)$$

To find a spatial and local behavior of the medium, the elastic analogy associated to the modified general relativity suggests the existence of deformations following the direction *N* (*N* = z, the direction of propagation of the gravitationnal wave) with Poisson's ratios $\nu_{NT} = \nu_{NL} \neq 0$ associated with equivalent normal stresses $\sigma_N$, *(L, T, N)* = (x, y, z).

## 5.4. Concerning the analogy between geometric torsion and crystal-plasticity of the equivalent elastic medium

This notion of a complementary vector to close a path on a surface, related to geometric torsion in the Einstein-Cartan theory, is also present in crystal plasticity **[40]** when, at the atomic level, local plasticification occurs **[56].** There are 2 types of defects: screw dislocations (Figure 15) and edge dislocations (Figure 14a), sliding by shear effect, and disinclination by forced rotations (Figure 14b).

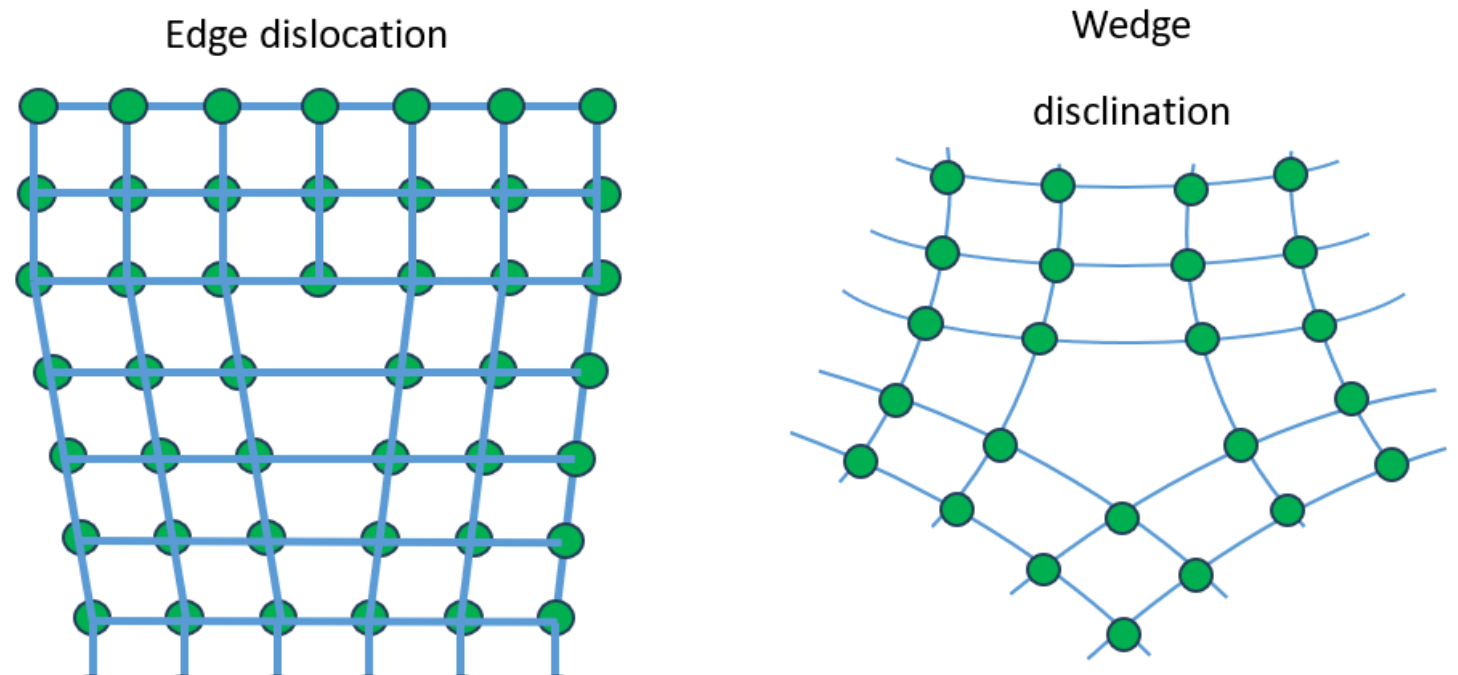

**Figure 14: Example of defects: Edge dislocation and disclination –**

In both cases, there is a discontinuity in the network. Mathematically, when we make a path through this dislocation, we must use a closure vector called the Burgers vector, which is equivalent to the geometric torsion explained above (Fig. 15) and formula (75).

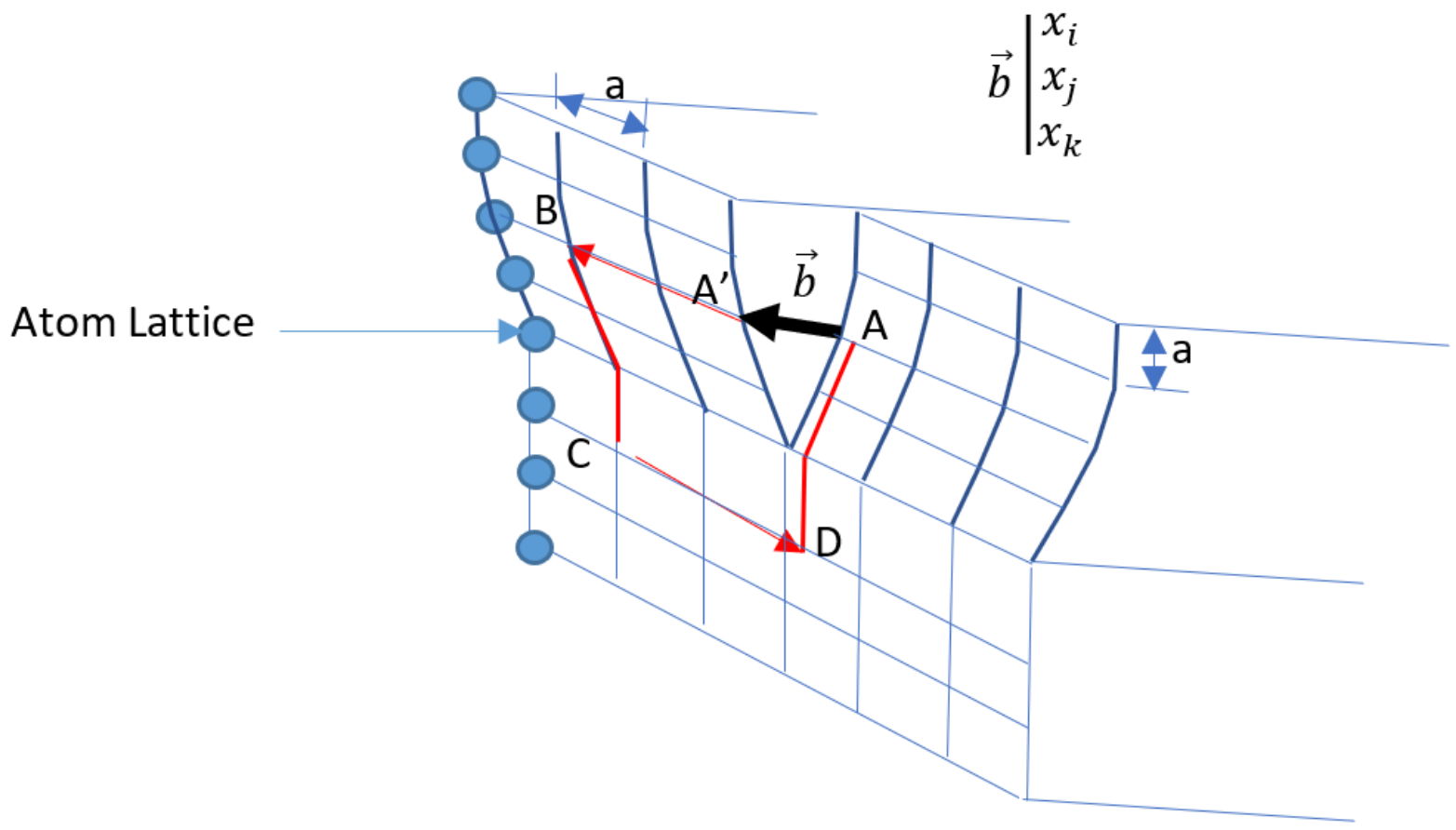

**Figure 15: Screw dislocation - Visualization of the Burgers vector $\vec{b}$ , the points A and A' –**

By writing the path $ABCD$ in Figure 15, we can see that mathematically the equivalent Burgers vector is expressed as follows **[56]**:

$$db^{\mu} = -\Gamma_{\nu\lambda}^{\ \ \mu} dA^{\nu\lambda} \qquad (39)$$

$dA^{\nu\lambda}$ being antisymmetric, the symmetric part of the affine bond is excluded, and only the antisymmetric part exists, which is written $\Gamma_{[\nu\lambda]}^{\ \ \mu}$. So, the equivalent Burgers vector is written:

$$db^{\mu} = -\Gamma^{\ \ \mu}_{[\nu\lambda]} dA^{\nu\lambda} \quad (40)$$

The bridge is then made with geometric twisting:

$$-T^{\lambda}_{\mu\nu} = -2\Gamma^{\lambda}_{[\mu\nu]} \quad (41)$$

By defining:

$$-[\nu\lambda] = \frac{1}{2}(\lambda\nu - \nu\lambda) \quad (42)$$

The theory of defects in crystal plasticity is therefore a mirror of geometric torsion in general relativity **[56]**. The analogy with two distinct mechanical concepts-are - elasticity and perfect plasticity - are not contradictory here. Indeed, the bridge between the theory of defects and the Einstein-Cartan theory with torsion, as explained above, will imply in the analogy of the elastic medium that planes of space could successively slide locally as a defect during the passage of a gravitational wave, as shown in Figure 16 below. This is compatible with a possible shear modulus of the medium. There would be no propagation of the torsion as such. Two alternatives are possible: either the space would plasticize locally by shear/distortion, or the analogy would reach its limit in this example.

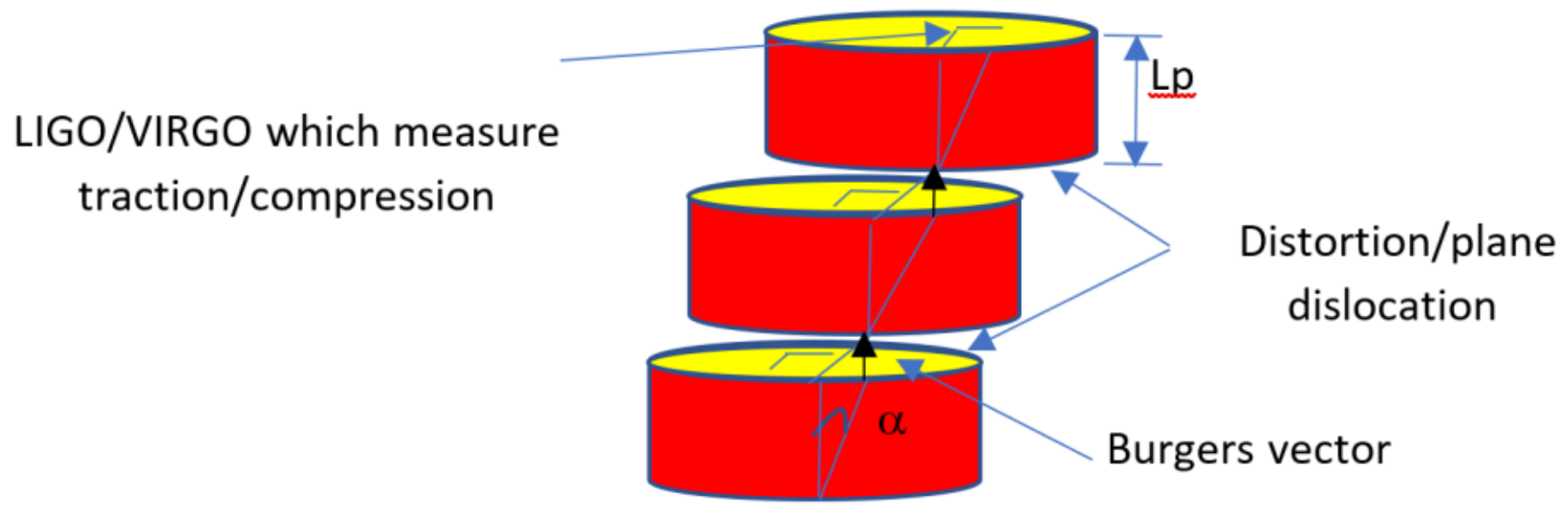

**Figure 16: Visualization of possible local dislocations between the different layers of the structure of the fabric of spacetime (analogy) during the passage of a gravitational wave –**

It is known **[36]**, **[37],** that the modified Einstein-Cartan theory of general relativity with geometric torsion comprises two field equations, one equivalent to that developed by Einstein (40) and another corresponding to spins of space (41). It has been shown by the authors in **[36] [37]** that the mathematical formalism associated with this geometric torsion via the Burgers vector (39) to (42) is similar to that in crystal plasticity **[63].**

It is shown via **[38]** that polarizations complementary to those predicted in classical general relativity and $(A^+, A^\times)$ appear when we consider this geometric torsion.

We know from **[9]** that the linearized Einstein equation leads to the squeeze (10) (equivalent to (11)), constituting a bridge that allows us to read polarizations in the elastic domain $A^+$ and $A^\times$ as components of a strain tensor. This approach has been illustrated in two concrete cases of space twisting in **[11]** and **[12]**.

However,–we have not yet shown that the second spin equation in the Einstein-Cartan theory corresponds to another equation of correspondence between polarizations and deformations in the plastic domain. Is there therefore a transitional formalism equivalent to equation (11) in plasticity associated with defect theory **[36]** and associated complementary polarizations **[9]** and **[11]**? In other words, in plasticity (corresponding to geometric torsion according to **[36]**), can complementary polarizations also be effectively read as components of a deformation tensor of a 4-dimensional space (and no longer only as deformations in successive planes independent of each other as shown in **[9]** and **[11]**) ? The bridge between the components of the polarization tensors of gravitational waves in the case of modified general relativity (Einstein-Cartan) and, by analogy, the deformations linked with an associated elasto-plastic medium was made in **[47]**.

The authors thus considered a gravitational wave as a defect (propagation of a Burgers vector) propagating in an equivalent solid medium. The result of their study is again a compression component H in space and shear and distortion components $\pm\sqrt{2}a_i$, as described in expression (43) in 4 dimensions and (44) in 3 dimensions.

$$\varepsilon_{\mu\nu} = \frac{1}{2}\begin{pmatrix} H & -\sqrt{2}a_1 & -\sqrt{2}a_2 & H \\ -\sqrt{2}a_1 & 0 & 0 & -\sqrt{2}a_1 \\ -\sqrt{2}a_2 & 0 & 0 & -\sqrt{2}a_2 \\ H & -\sqrt{2}a_1 & -\sqrt{2}a_2 & H \end{pmatrix} \quad (43)$$

They also cite the existence of what we called a torsion wave in **[11]**, which they refer to as a J-dependent giratonic wave (a spin of space), in which the function $J$ is related to the spinning nature of the gyratons **[47]**:

$$\varepsilon^{(i)(j)} = \frac{1}{2}\begin{pmatrix} 0 & 0 & \frac{\sqrt{2}J}{\rho A}\sin\Phi \\ 0 & 0 & -\frac{\sqrt{2}J}{\rho A}\cos\Phi \\ \frac{\sqrt{2}J}{\rho A}\sin\Phi & -\frac{\sqrt{2}J}{\rho A}\cos\Phi & H/A \end{pmatrix} \quad (44)$$

Note that in this study [**47**], the authors placed themselves within the framework of nonlinear plane gravitational waves, or parallel-propagating plane front waves (P-P waves) **[52].** Indeed, as this study considers the parallelism between crystal plasticity, the implementation of which characterizes a plasticization of the medium by rearrangement of atoms (and therefore a non-linearity between deformations and stresses) **[36],** and Einstein Cartan's nonlinear modified general relativity associated with this theory **[36]** to **[38]**, it makes sense to move away from the realm of traditional elastic waves to consider nonlinear plane waves known as PPs.

Note also that the result they obtain (expressions 43 and 44) for a gravitational wave propagating in the direction $z$ is connected to polarizations $A^{+}$ and $A^{\times}$ gravitational waves in classical general relativity via the expressions (45) and (46) of **[52]**:

$$H = A_{+}(u)(x^2 - y^2) \; (45)$$

$$H = A_{\times}(u)xy \; (46)$$

Recently, in **[48]** to **[51],** the authors consider a shape memory of space (i.e. a certain residual plasticity of this medium).

The null values in the above two tensors (43) and (44) result from the fact that the authors focused on the geometric torsion part associated with the Burgers vector itself, which is associated with crystal plasticity **[36]**, i.e., the second equation of the Einstein-Cartan theory that presents an analogy with this theory. Thus, as in **[38]** where complementary polarizations appear because of this geometric torsion in the components $(zz)$, $(zx)$, $(zy)$ and $(zt)$ in 4 dimensions, in **[47]**, complementary deformations appear according to these same components and only for them. However, this publication **[47]** shows that, unlike the classical equation of general relativity where the bridge between polarization and deformation is direct via expression (11), this time, for the torsion component, the correspondence between the components of the polarization tensors and the deformation tensors in 4 and 3 dimensions is no longer direct. The publication **[47]** explains mathematically how to make this transition.

On the other hand, our paradigm for reading the components of polarizations as components of a strain tensor remains the same.

### *5.5. Concerning the convergence of the different models regarding possible polarizations of gravitational waves in the direction of their propagation*

In order to analyze the convergence between the different approaches resulting from general relativity, we considere the following:

Concerning the 1st case: (second-order linearization of $h_{\mu\nu}$ **[34]**, **[35]**, the Einstein-Cartan theory of geometric torsion **[36], [37]**, the theories of modified general relativity **[39]**, **[44]**) and approaches derived from the analogy of elastic space, we find polarizations complementary to the two classics $A^{+}$ and $A^{\times}$. Concerning the 2nd case: (strong anisotropy of the medium **[9]**, crystal plasticity **[56]** (junction of transverse planes by successive plastic slips)), we can see by virtue of the analogy between the polarizations of gravitational waves and the deformations of the equivalent elastic medium **[9]**, **[11]**, **[12]**, longitudinal spatial deformations complementing the transverse deformations.

It should be stressed that this convergence aims to continue optimizing the performance of interferometric sensors, to add arms to have a triangular measurement system (Future LISA or Einstein telescope) to possibly detect them and thus settle the question.

### *5.6. Concerning the extreme smallness of the deformations/polarizations in the longitudinal direction of gravitational wave propagation (if they really exist)*

If such longitudinal deformations/polarizations exist, the measurement of the gap between theoretical and real gravitational waves (Figure 5) seems to indicate that they are much smaller than those already measured by the current LIGO/VIRGO interferometers **[41] [3] [4]**. The second-order study of $h_{\mu\nu}$ in gravito-electromagnetism is more nuanced, the intensity of the out-of-plane deformations depends on the frequency of the gravitational wave **[34] [35].**

### *5.7. Concerning the importance of excellent coordination of interferometers on Earth complemented by future LISA type interferometers or the future Einstein telescope or multiple pulsars*

Publications **[41]** to **[46]** show that many research teams are currently working to theorize and measure these potential complementary polarizations of gravitational waves, including in the direction of gravitational wave propagation. In **[44]**, the author explains how these different possible polarizations could be studied by an even more efficient interconnection of the different interferometers on Earth.

## 6. Conclusion

*Concerning the convergence between the analogy of space as an elastic medium and the results of general relativity*

The analogy of continuum mechanics with general reativity in the weak field is interesting because it allows us to better understand certain aspects of the latter. The metric perturbation tensor is assimilated with twice the strain tensor **[9]**, we notice the similarity between the stress tensor and the energy-momentum tensor **[9]**, **[11]**. It is possible to consider the parameterizable constant κ as a function of the mechanical characteristics of space **[11]**. Then, Einstein's equation appears similar to Hooke's law **[9]**, **[11]**, **[12]**, and finally, gravitational waves are similar to medium shear waves, **[9]**, **[11]**,...). It also provides a new illustration of the origin of the two polarisations instead of one or 6 (pure twisting of an elastic medium) **[11]**. Finally, it suggests distortion (lateral displacement of laser beams of interferometers).

*Concerning the divergence between the analogy of space as an elastic medium and the results of general relativity*

The analogy of elasticity theory with general relativity in a weak field also reaches its limits and raises questions about its real representativeness, given the extremely high intensity of the associated Young's moduli **[9]** to **[11]**, the value of the Poisson's ratios of 1 in the interferometer plane and 0 out of plane, leading to a strong anisotropy of the medium in any point M and during wave propagation, contrary to the fundamental hypothesis of the homogeneity of the cosmic medium.

*Concerning the convergence between current research and what the analogy of space as an elastic medium with the results of general relativity suggests*

The discrepancy between the measured and theoretical curves of space deformations during the passage of a gravitational wave (Figure 5) suggests that there is still room for improvement in the general relativity. However, this improvement is a priori extremely small, as this gap is very small **[3]** and **[4]**.

Research avenues to try to complete this general relativity concern, in particular, potential complementary polarizations of gravitational waves **[38]**, **[39]** and **[44]**.

The modified elastic analogy of general relativity with Einstein-Cartan geometric torsion connected to the theory of defects **[36]** to **[38]**, as well as the approach of general relativity developed in the second order **[34]**, **[35]**, involve additional deformations in the direction of propagation of gravitational waves. Such approaches make it possible to find a coherent spatial behavior and therefore a little less anisotropy of the elastic space associated in our analogy. The geometric torsion, as developped for example by Einstein-Cartan theory can be assimilated at a plasticity between each sucessive planes via the analogy of the cristallography and defect theory. On this basis, a 3D equivalent anisotropic space medium become possible (polarisation in 3 directions and not only on transverse plane, coupled by analogy with strains in 3 dimensions and not only in 2 dimensions) **[34]**, **[35]**. It should also be noted that these additional corrections are extremely small **[3] [4]** and are therefore consistent with what is revealed by the measurements of the deviation between the measured deformation curves associated with gravitational waves and the same deformation curves from unmodified general relativity **[3]**, **[4]**. According to our paradigm associating these deformations with complementary polarizations, this implies detecting them in order to validate or invalidate these deformations associated with these complementary polarizations.

The multiplicity of recent publications on these potential complementary polarizations in relativistic physics on the one hand **[41]** to **[46]** and on how to measure them on the other hand indicates (future LISA and Einstein telescope) that this topic is a key research point today. Time will tell whether or not this precise point is a point of convergence between the analogy of the elastic medium and general relativity in the weak field.